\documentclass[11pt,a4paper,twocolumn]{article}
\usepackage[margin=1.6cm,bottom=1in]{geometry}

\usepackage{multicol}
\usepackage{abstract}
\usepackage{enumerate}

\usepackage{graphicx}
\DeclareGraphicsExtensions{.png,.jpg,.pdf}
\usepackage{subcaption}

\usepackage{amsmath}
\usepackage{amssymb}
\numberwithin{equation}{section}

\newif\ifanglestyle           \anglestylefalse
\newif\ifbracestyle           \bracestylefalse
\newif\ifbracketstyle         \bracketstylefalse
\newif\ifparenthesisstyle     \parenthesisstylefalse
\newif\ifcalligraphicstyle    \calligraphicstylefalse
\newif\ifnoncalligraphicstyle \noncalligraphicstylefalse

\RequirePackage{amsbsy}
\RequirePackage{amsopn}
\RequirePackage{amsfonts}

\RequirePackage{relsize}
\RequirePackage{exscale}

\RequirePackage{ifthen}

\usepackage{algorithm}
\usepackage{algorithmic}

\usepackage{caption}
\usepackage{orcidlink}

\DeclareMathOperator*{\argmin}{arg\, min}
\DeclareMathOperator*{\argmax}{arg\, max}
\newcommand{\R}{\mathbb{R}}

\usepackage{ifthenx}
\newboolean{IF}
\setboolean{IF}{true}

\DeclareMathOperator{\e}{\mathbb{E}}
\DeclareMathOperator{\loss}{\mathcal{L}}
\DeclareMathOperator{\sspace}{\mathcal{S}}

\newcommand{\prast}{p(a_t\vert s_t)}
\newcommand{\prssa}{p(s_{t+1} \vert s_t, a_t )}

\newcommand*{\RNum}[1]{\uppercase\expandafter{\romannumeral #1\relax}}

\renewenvironment{abstract}
  {\begin{center}
    \bfseries Abstract\vspace{-1em}\vspace{0.5em}
  \end{center}
  \list{}{%
    \setlength{\leftmargin}{0pt}
    \setlength{\rightmargin}{0pt}
  }%
  \item\relax}
  {\endlist}

\title{\vspace{1pt}Exploiting Distributional Value Functions for \\Financial Market Valuation, 
Enhanced Feature Creation \\and Improvement of Trading Algorithms\vspace{1pt}}

\author{Colin David Grab\thanks{\textbf{DISCLAIMER}: The presented research was conducted and implemented independently without any affiliation to any company or support thereof. All contents and opinions expressed in this document are solely those of the author and do not represent the view of any  company or institution. The author disclaims any representation or warranty for its accuracy or completeness. It is not intended as investment advice, services or an endorsement of any specific investment or strategy. Copyright \copyright \,2024, C. Grab \orcidlink{0009-0000-4120-8957}. All rights reserved.} \\  \vspace{1pt} {\small cdg.paper@gmail.com} }
\date{May 2024}

\begin{document}
\bibliographystyle{abbrv}   

\twocolumn[
\maketitle
\begin{abstract}
\normalsize
\textit{Many research studies of reinforcement learning applied to financial markets have been published in recent years. 
While they predominantly concentrate on finding optimal behaviours or trading rules, it is worth to take a step back and realize that the reinforcement learning returns $G_t$ and state value functions themselves are of interest and play a pivotal role when it comes to the evaluation of assets. Thus, in this paper instead of focussing on the more complex task of finding optimal decision rules, the power of distributional state value functions is 
for the first time studied and applied in the context of financial market valuation and machine learning based trading algorithms. 
Although the values of $G_t$, representing a time weighted average of future market changes, cannot be known explicitly ahead of time, accurate and trustworthy estimates of their distributions and expected values provide a competitive edge leading to better informed decisions and more optimal behaviour.
Herein, ideas from predictive knowledge and deep reinforcement learning are combined to introduce a novel family of models called CDG-Model, resulting in a highly flexible framework in the context of financial markets and intuitive approach with minimal assumptions regarding underlying distributions. The models allow seamless integration of typical financial modelling pitfalls like transaction costs, slippage and other possible costs or benefits into the model calculation. They can be applied to any kind of trading strategy or asset class. The frameworks introduced provide concrete business value through their potential in market valuation of single assets and portfolios, in the comparison of strategies as well as in the improvement of market timing. In addition, they can positively impact the performance and enhance the learning process of existing or new trading algorithms.
They are of interest from a scientific point-of-view and open up multiple areas of future research.
Initial implementations and tests were performed on real market data. While the results are promising, applying a robust statistical framework to evaluate the models in general remains a challenge and further investigations are needed.}
\end{abstract}
{\small \textbf{Keywords:} Quantitative Finance, Machine Learning in Finance, Deep Reinforcement Learning, Predictive Knowledge, Portfolio Optimization, Algorithmic Trading, Financial Markets, AI}
\vspace{10pt}
]
\saythanks
\section{Introduction}
A major challenge of models for financial markets is to find a suitable set of input data. 
The incorporation of transaction costs and slippage into financial models presents another difficulty, especially when applying theoretical models, which are often based on restrictive assumptions, to financial markets. While modern machine learning algorithms might be less dependent on assumptions and modelling choices, they often are less interpretable and suffer from the challenging need to learn a suitable representation of the data, while at the same time they need to learn some optimal behaviour or prediction function. The presented model framework herein addresses and tries to alleviate each of these challenges.

This paper introduces a novel model family of algorithms applicable in the context of financial market valuation, feature creation and trading algorithms based on machine learning.

The main advantage of this family of models is that they do not rely on specific distributional assumptions and that they have favourable properties regarding uniqueness and existence in theory. They are not restricted to specific assets or machine learning algorithms, but offer a wide range of choice to tailor them to relevant cases of interest.\\

This model family combines ideas from predictive knowledge with concepts and approaches from deep reinforcement learning and translates them to financial markets. Specifically, it investigates if the concept of value functions and distributional value functions from deep reinforcement learning in combination with ideas from predictive knowledge can be applied to evaluate financial assets or portfolios. 

The advantage of using a value function based approach over some point estimation process to predict future prices or some mathematical model relying on very specific assumptions, derives from the fundamentals of value functions and their well established mathematical properties. 
In comparison to other approaches, estimating value functions requires minimal assumptions\footnote{Simplified, the conditional probabilities of the process need to be stationary.} regarding the distribution, and if satisfied, the value functions exist and are unique \cite{Sutton_Barto}.  
Several unbiased estimation methods for state value functions with convergence properties exist. 
In addition, their definition is simple to understand and intuitive due to its strong similarity to discounted cash flow methods. 

Predictive knowledge or general knowledge focuses on the intuition that models which are trained to solve different problems with the same input data are forced to learn a relevant representation of "the world". These representations not only focus on one specific aspect, but also need to contain information relevant to understand and solve different dynamics and problems within the same environment \cite{Bengio_2013},\cite{Littman_2002},\cite{Sutton_2011},\cite{White_2015}.
Translating this concept to machine learning has led to applications where models are built and trained not only to solve one main objective, but to simultaneously solve multiple other auxiliary objectives. 
The scope of such additional tasks can vary widely from objectives very similar to the main problem, objectives concerning sub- or related problems, but also objectives that focus on more technical aspects of the model themselves, like for example maximizing activations of layers in a neural network.
Intuitively, a model that \textit{understands the world better in general} will have learnt a more informative representation of the data and is better equipped to solve the main objective. \\

The proposed models aim to support the creation of more information dense features by learning in parallel multiple quantities relevant in financial markets. Thereby they ideally learn a representation that better captures underlying mechanisms and allows for greater understanding of financial markets. These models and their learned feature representation can potentially be used on a stand alone basis as a pure valuation approach or as a supporting and stabilizing component of machine learning based trading frameworks.\\

Besides the theoretical advantages, the approach allows to seamlessly incorporate transaction costs, slippage and any other relevant costs or benefits directly into the calculation of the model environment. If available, even real world transaction data from past activities can be exploited. Moreover, additional performance measures can be derived by choosing an appropriate definition of the rewards as introduced below.

All combined, this leads to a very flexible framework with desirable properties and minimal assumptions to investigate financial markets. The framework is presumably best suited for trading frequencies in the minutes range or even high-frequency range.\\

For completeness and comprehensibility, the paper is structured as follows:
First, thorough fundamental theory and frameworks from reinforcement learning are reviewed. These include both the estimation of value functions as well as a way of estimating their distributional analogue.
Secondly, to better illustrate the approach, the model is introduced both as a simpler variant focusing only on estimating an expected value as well as a more involved distributional version. Subsequently, the setup for these type of models in the context of financial markets and different components of the approach are discussed.

Finally, a first set of results are presented from studying the performance of model variants on real world financial data, specifically applied on prices for stocks, indices as well as crypto currencies. These results are just a first indication, however the implementations serve as proof of concept of a successful functional approach. They strongly motivate further investigation of these kind of models and their potential benefits.

\section{Theoretical Foundations}
The necessary concepts of reinforcement learning to be applied below are hereby recalled. The theoretical foundations are split in three main topics: fundamental theory, stabilization of function approximation and distributional reinforcement learning.

The informed reader can skip directly to the introduction of the approach in Section \ref{sec:approach}. 

\subsection{Background on Reinforcement Learning}\label{rl_theory_fund}
Reinforcement learning is a machine learning approach that is used to learn behaviours in scenarios in which an agent interacts with an environment around him. 
At every step in time, this agent decides how to act based on the state of his surroundings. Then, depending on his action and the reaction of the environment, the agent receives positive or negative feedback from the environment in form of some reward. 
The overall goal of the agent is to behave such that he accumulates as many positive rewards as possible. 
Reinforcement learning provides the agent the capabilities to adjust his behaviour based on this feedback. 
Ideally, this enables the agent to learn optimal behaviour by encouraging decisions that receive positive feedback, while discouraging decisions that lead to negative feedback. 
In contrast to the other machine learning categories, supervised learning and unsupervised learning, in reinforcement learning the algorithm learns by itself solely from the interaction with the environment and its feedback. The optimal behaviour and rules of the environment do not need to be known in advance.\\
This setting of an agent interacting with an environment is formalized and modelled as a Markov Decision Process (MDP). In a MDP, an agent interacts with an environment at consecutive, discrete time steps $t=0,1,\dotso T$. 
At every time step $t$, the agent observes a representation of the environment, representing the so called "state" at time $t$. The state is denoted by the random variable $S_t \in \sspace$, with $\sspace$ being the space of all possible states. After observing the state, the agent chooses an action $A_t \in \mathcal{A}(s)$, where $\mathcal{A}(s)$ denotes the space of all possible actions for some realized state $s$, i.e. $S_t = s$. 
As a consequence of this action $A_t$, the agent receives a numerical reward $R_{t+1} \in \mathcal{R} \in \mathbb{R}$ and transitions to a new state $S_{t+1}$. 
In an MDP, the Markov assumption implies that the probabilities\footnote{$P(S_{t+1},R_{t+1}|S_t,A_t,S_{t-1},A_{t-1},S_{t-2},A_{t-2}\dotso)$ $=P(S_{t+1},R_{t+1}|S_t,A_t)$.} of the next state and reward depend only on the current state $S$ and action $A$. 
For a finite MDP, the conditional probability distribution of the random variables $s' \in \sspace, r \in \mathbb{R},$ given the current realizations $s \in \sspace, a \in \mathcal{A}$ is defined as:
$p(s',r|s,a) := P \left(S_{t+1} = s', R_{t+1} = r | S_{t} = s, A_{t} =a \right)$, where $\sum_{s'\in \sspace} \sum_{r \in \mathcal{R}} p(s',r|s,a) =1,$ for all $s \in \sspace, a \in \mathcal{A}(s)$. 
From these probabilities, if known, one can compute all relevant probabilities \cite{Sutton_Barto}.

Since the overall goal is to learn how to act optimally, there needs to be a quantity that allows to measure and compare the performance of an agent. This quantity is called the return\footnote{Because a return is also a commonly used quantity in finance, if not clear from the context, return will now exclusively be used to reference the return in the context of reinforcement learning theory.} $G_t$ at time $t$ and is defined\footnote{This definition includes the possibility of $T=\infty$ or $\gamma = 1$ as long as not both hold at the same time \cite{Sutton_2011}.} as the time weighted sum of all future rewards an agent accumulates over time
\begin{equation}
G_t = \sum_{k=t+1}^T \gamma^{k-t-1}r_k.
\end{equation}
Here $\gamma \in [0,1]$ is a discount factor and can be interpreted as a constant probability of termination given by $(1-\gamma)$ \cite{Sutton_Barto},\cite{Sutton_2011}, or similar to the use of discount factors in economics, as a weighting of future pay-offs that reflects the time value of future rewards. Hence, from an economic perspective, a return $G_t$ has high similarity to the concept of a "present value". Another possible way of thinking about the return $G_t$, is to think of it as a weighted moving average over future rewards, where the importance of the future depends on the discount factor $\gamma$.

The decision rule of an agent, i.e. the specification which actions to take in which state, is called the policy $\pi(\cdot)$. A policy can be deterministic or stochastic. A deterministic policy is a mapping from states to actions $ \pi(s) \mapsto a $ and a stochastic policy $\pi(a|s)$ is a mapping from states to conditional probabilities of the actions given the state, i.e. $\pi(a|s) = P(A=a|S=s),$ where $\int_{a \in \mathcal{A}} p(a|s) =1$.

When an agent interacts with its environment over time this generates a trajectory 
\begin{equation}
\tau = (s_0,a_0,r_1,s_1,a_1,r_2,s_2,a_2, \dotso, r_T,s_T ). \nonumber
\end{equation}
The probability of this trajectory is defined as
\begin{equation}
 p(\tau) = p(s_1) \prod_{t=1}^T \prast \prssa.  \label{eq:TrajDeb}
\end{equation}

The expected return at time $t$ over the trajectory introduced by following a policy $\pi,$ is defined as the state value function $V_\pi(s):$ 
\begin{align}
V_\pi(s) &= \e_\pi \left[G_t | S_t = s \right]  \\
&= \e_\pi\left[\sum_{k=t+1}^T \gamma^{k-t-1}r_k | S_t = s \right] \nonumber \\
&= \e_\pi\left[ r_{t+1} + \sum_{k=t+2}^T \gamma^{k-t-1}r_k | S_t = s\right]. \nonumber
\end{align}

Here, a subscript $\pi$ is used to represent the dependency\footnote{Dependency on the probabilities over which the expectation is calculated: $a \sim \pi(a|s),s' \sim p(s'|s,a),r \sim p(r|a,s').$} of the random variables of the trajectory on the policy followed to generate the trajectory. Whenever it is clear from the context, the subscript $\pi$ will be dropped in the following to simplify notation.

A simpler setting can be formalized similarly as a Markov Reward Process (MRP) \cite{Silver_Hasselt_2017}. In a MRP no decisions need to be taken, and rewards are defined by some process $\mathbf{p}$, that depends on the state and some randomness $\alpha$, i.e. $S_{t+1},R_{t+1} \sim \mathbf{p}(s,\alpha)$. The same basic properties and a state value function $V_\mathbf{p}(s)$ depending on the process $\mathbf{p}$ are also defined in the context of an MRP.
In both MDP and MRP, the state value functions $V_\pi(s),V_{\mathbf{p}}(s)$ follow the inherent recursive relationship 
\begin{align}
V(s) = \e \left[ r_{t+1} + \gamma V(s') | S_t = s \right] \label{Eq_Bellman},
\end{align}
that holds for all $s \in \sspace$ and is known as the Bellman equation \cite{Bellman}.
The value function $V(\cdot)$ is the unique solution to the Bellman equation. The Bellman equation defines a contraction operator, which when repeatedly applied converges to the real value function. 
Most importantly, the existence and uniqueness of the value function in an MDP or MRP is guaranteed, if either $\gamma <1$ or the termination of the trajectory is guaranteed by $T$ being finite \cite{Sutton_Barto},\cite{Bellman},\cite{Donoghue_2016}.

In small, finite cases these value functions can be calculated using tabular approaches and for known transition probabilities the optimal solutions can be calculated explicitly. However, in most real-world application, such as in the case of financial markets, the state space and action space are too complex and the transition probabilities and true state space are not known. 
In these cases, function approximation methods with some parametrized function $f(\cdot; \theta)$ are used. This $f(\cdot;\theta)$ maps some input to an estimate that depends on the functions parameters $\theta$, often called weights. 
Gradient methods can be used to adjust the parameters of the function to optimize some performance measure or loss function. 

Usually, reinforcement learning is used to learn an optimal policy $\pi^*(S)$ that maximizes the expected return, i.e. $\pi^*(S) = \argmax_{\pi} \e_{a\sim \pi} [G_t|S_t=s],$ with the use of state value functions and other concepts\footnote{Such as state-action value functions known as Q-functions \cite{Watkins_1989}, advantage functions \cite{Sutton_1999},\cite{Baird_1994}, or policy improvement theorems and actor-critic algorithms etc. \cite{Silver_Lever_2014},\cite{Sutton_1999},\cite{Williams_1992},\cite{Konda_2000},\cite{Lilicrap_2015}.} which are not of relevance here.

\subsubsection{Learning State Value Functions}
The process of function approximation to learn the state value function is described in the following.

Let $v(s)$ denote the true, but unknown value function and $\hat{v}(s) :=v(s;\theta) = v_\theta(s)$ a parametrized function to approximate the true value function. Since the objective is to find a parametrized function as close to the true value function as possible, i.e. $\hat{v}(s) \approx v(s)$, the usual approach is to minimize the mean squared difference between the estimated value and the true value over all states:
\begin{equation}
\min \sum_{s\in \sspace} \mu(s)[v(s) - \hat{v}(s;\theta)]^2,
\end{equation}
where $\mu(s)$ denotes the state probabilities.

However, both the true value function $v(s)$ and the state probabilities $\mu(s)$ are not known in advance and realistically the real state space is unknown and too large to be summed over. 
To remedy this, two approaches need to be taken. 
First, one must find some feasible estimate to substitute for $v(s)$ as a target. 
Secondly, one needs to collect sufficient samples by simulation or real word interaction data and approximate the true error with a sample estimate by averaging the realized error over many samples.

One popular approach is to utilize the recursive relationship\footnote{The difference between the quantities is often called the temporal difference error $\delta_t$ (TD-error, \cite{Sutton_Barto}),  \quad \quad \quad \quad \quad \quad \quad
i.e. $\delta_t = R_{t+1}+\gamma V(S_{t+1}) - V(S_t).$}
defined in the Bellman equation and to replace ${v}(s)$ in the loss by a target estimate
\begin{equation}
{v}(s) = r_{t+1} + \gamma \hat{v}(s';\theta).
\end{equation}

Using the recursive relationship and the uniqueness of the value function, the repeated application of the Bellman operator\footnote{i.e. using the Bellman equation (Eq. \ref{Eq_Bellman}) to define a target.} contracts to the true value function. 
Put differently, if one can find a parametrized function, such that the relationship $\hat{v}(s;\theta) = r_{t+1} + \gamma \hat{v}(s';\theta)$ holds for all $s \in \sspace$, then the approximated function $\hat{v}(\cdot)$ must be equal to the true state value function due to its uniqueness property. 

Mathematically, the goal is to find the function parameters $\theta$ that minimize the objective:
\begin{equation}
\theta^* = \argmin_\theta \sum_{s\in \sspace} \mu(s)[r_{t+1} + \gamma \hat{v}(s';\theta) - \hat{v}(s;\theta)]^2. \nonumber
\end{equation}

In a real life scenario such as financial markets, one can not compute the state value for all states, but must generate sample transitions $(s,r,s')$ and estimate the loss over batches of sampled transitions instead of $\sum_{s\in \sspace}$.

Another option for a target in the loss estimation is a recursive substitution of the value function in the right-hand side of the Bellman equation (\textit{Eq.} \ref{Eq_Bellman}). This leads to the following form
\begin{align}
V(S_t) &= \mathbb{E} \left[ r_{t+1} + \gamma V(S_{t+1}) \mid S_t = s \right] \nonumber \\
& \begin{aligned}
&= \mathbb{E} \Bigl[ r_{t+1} + \gamma r_{t+2} + \gamma^2 r_{t+3} + \\
&\qquad \cdots + \gamma^{n-1} r_{t+n}+ \gamma^n V(S_{t+n}) \Bigr], \nonumber
\end{aligned}
\end{align}
which can be translated to construct n-step targets to use in the estimation as $r_{t+1} +  \gamma r_{t+2} + \dotso + \gamma^{n-1}r_{t+n} + \gamma^n V(S_{t+n}).$

For better readability below, we additionally denote 
\begin{align}
r_{t:t+n-1} &=r_{t+1} +  \gamma r_{t+2} + \dotso + \gamma^{n-1} r_{t+n} \label{eq:r_in_nsteps} \\
&= \sum_{k=t+1}^{t+n} \gamma^{k-t-1}r_k. \nonumber 
\end{align}
Applying n-step targets provides an estimate with lower variance trading off sample-efficiency\footnote{Note that the choice of number of steps to include, yields an n-step variant somewhere between $n=1$ and $n=T$ (a Monte-Carlo method) \cite{Sutton_Barto},\cite{Watkins_1989}.} \cite{Sutton_1988},\cite{Sutton_Barto}.

\subsection{Stabilizing the Approximation}\label{rl_theory:approx}
Unfortunately, many of the convergence guarantees from the finite setting fail in complex state spaces, especially when using deep non-linear function approximates.  Convergence and stability of the learning process is not guaranteed and convergence can sometimes reach suboptimal levels or is only given for very specific parametrizations. 

One of the major reasons for unstable behaviour of deep non-linear function approximators in reinforcement learning is that in contrast to most supervised learning algorithms, which work under the premise of independently and identically distributed samples, reinforcement learning algorithms work with samples and targets from sequences of observations which exhibit strong temporal correlations as well as non-stationary distributions \cite{Mnih_2013},\cite{Mnih_2015}.
While in supervised learning the learning targets are fixed and known, the learning targets in value function estimation are themselves output of the function and their distributions change constantly.

For a more detailed discussion about problems such as divergence or instability arising with function approximators consult for example \cite{Tsitsiklis_1997},\cite{Thrun_1993},\cite{VanHasselt_2018}. \\

In order to remedy this, several approaches have been developed and introduced in the space of deep reinforcement learning. These approaches aim to reduce the variance in the estimates, try to facilitate convergence of the algorithms, attempt to increase their robustness and data efficiency, all while trying to keep them un- or very low-biased. 
The authors in \cite{Mnih_2015} propose two adjustments to mitigate these problems, namely the use of experience replay to decorrelate sample transitions and the use of so called target networks to stabilize the targets distribution. 
Further developments of these two approaches have established themselves throughout the deep reinforcement literature and have been shown multiple times to increase both performance and robustness of the learning process\footnote{Ablation studies investigating the effects of different algorithmic adjustments and their contribution to increased performance and stability of the algorithms are presented in \cite{Hessel_2018c} and indicate that prioritizing samples is the most important contribution to increased performance and stability of the algorithms.}.

Some of these approaches, specifically the use of prioritized replay buffers, target networks with soft-updates as well as using n-steps estimates are integrated in the implementation of the models at hand and shortly recalled below.

\subsubsection{Experience Replay Buffer}
The first approach uses replay buffers, reintroduced as experience replay in \cite{Mnih_2013} and \cite{Mnih_2015}. In this technique, samples of transitions, i.e. experiences $e_t = (s_t,a_t,r_t,s_{t+1}),$ are collected by interaction of the agent with the environment and stored into a data set called a replay buffer, denoted by $\mathcal{B} = \{e_1,e_2,\dotso, e_t\}$.
During learning, batches of transitions are sampled from the buffer $\mathcal{B}$ and gradients (to update function parameters) with respect to the relevant loss are estimated based on these samples. 
The central idea is to break serial correlations because samples from a trajectory naturally are highly correlated. By sampling from a memory object with many samples, samples will originate from different trajectories and thus hopefully be less correlated and closer represent independently and identically distributed data as usually assumed in learning approaches. 

While in the initial version in \cite{Mnih_2013},\cite{Mnih_2015} transitions are sampled uniformly from a replay buffer, the approach was further extended in \cite{Schaul_2015b} to a version based on prioritized sampling of \cite{Moore_1993}. 

Prioritized experience replay assigns a scalar priority representing the importance\footnote{The reasoning is that some samples do not contain important knowledge, while other samples, e.g. such with large errors, contain more relevant information to adapt to.} of a sample for the learning process to each transition in the replay buffer. 
The authors in \cite{Schaul_2015b} define the priority $p_i$ of a sample $i$ proportional to the absolute value\footnote{$p_i = |\delta_t| + \epsilon$, where $\epsilon$ is some small constant added such that no sample will have a priority of zero.} of the temporal difference error of that sample for $i=1,\dotso , n_{\mathcal{B}},$ where $n_\mathcal{B}$ is the number of samples in the replay buffer.
Based on these priorities, the probability of each sample to be picked is calculated as 
$P(i) = \frac{p_i^\alpha}{\sum_k p_k^\alpha},$
where $\alpha$ is a hyper-parameter.
At every iteration during learning, the model samples a batch according to these probabilities from the replay buffer, i.e. $(s,a,r,s') \overset{P(i)}{\sim} \mathcal{B}$. Every time a transition is sampled from the replay buffer and used for updates, the priority of that sample is updated relative to the new error.
To correct for bias introduced by applying prioritized sampling, the authors in \cite{Schaul_2015b} introduce a variant of importance-sampling weights $w_i$ that are included into the update by multiplying the estimated loss for a sample $i$ by $w_i$ (see \textit{Eq.} \ref{eq:stabilized_loss}). The importance weights are given by
\begin{equation}
w_i = \left( \frac{1}{n_{\mathcal{B}} P(i)} \right)^\beta,
\end{equation}
where $\beta$ is a hyper-parameter controlling the correction size. There is an interaction between hyper-parameters $\alpha$ and $\beta$. Increasing $\alpha$ leads to higher prioritization, while increasing $\beta$ leads to more correction for introduced bias\footnote{The authors \cite{Schaul_2015b} recommend to scale the weights by the maximum weight and to anneal the hyper-parameter $\beta$ towards one.}.

\subsubsection{Target Networks}
Value approximation methods use the function approximation $v(S;\theta)$ for both the actual value as well as its target value. This is similar to "trying to hit a moving target" and promotes instability. 
To address the issue of the dependence and correlation between estimation and their targets, the authors in \cite{Mnih_2015} introduced a technique called target networks. 
A target network $v_{target}(S;\tilde{\theta})=v_{\tilde{\theta}}(S)$ is a copy of the approximation function $v(S;\theta)$ and is used to estimate the target. 
Applying target networks to estimate the target reduces the correlation and stabilizes the updates, thereby making the learning process more robust. The concept of target networks aligns the approximation problem closer with supervised learning. This is favourable because stable, robust approaches exist in supervised learning \cite{Lilicrap_2015}. 

Most algorithms now apply soft updates\footnote{As opposed to periodic updates of target parameters as in their introduction in \cite{Mnih_2015}.} as introduced in \cite{Lilicrap_2015} to the target parameters instead of periodically updating the parameters. Soft target parameter updates let the target parameters track the current parameters $\theta$ of the approximation function by using the following update rule:
\begin{equation}
\tilde{\theta} \leftarrow \tau \theta + (1-\tau) \tilde{\theta},
\end{equation}
where $\tau\in (0,1),$ usually $\tau \ll 1$, is some constant hyper-parameter determining the speed of how much the target network is lagging. \\

The combination of using transitions sampled from a prioritized replay buffer and soft-target functions leads to the following estimate of the loss over a batch consisting of $n_{batch}$ number of samples
\begin{equation}
\hat{\mathcal{L}} = \sum_{i=1}^{n_{batch}} w_i [r_{i} + \gamma \hat{v}(s'_{i};\tilde{\theta}) - \hat{v}(s_i;\theta)]^2,\label{eq:stabilized_loss}
\end{equation}
where $s_i,r_i,s_i',w_i$ correspond to the states, rewards, next states and importance weights of the $i$-th sampled transition in the batch.

\subsection{Distributional Reinforcement Learning}\label{sec:Distributional}
While state value function estimation focusses on the expectation of the return $G_t$, distributional methods aim to learn the distribution over the returns $G_t$, i.e. approaches to estimate $P(G_t|S_t).$
They are an analogous expansion of the state value functions from expected value to the complete distribution. The distributional view has been around for almost as long as the original Bellman equation \cite{Bellemare_2017}.
The following discussion is mainly based on the $C\text{-}51$ reinforcement learning algorithm introduced by the authors in \cite{Bellemare_2017}. Their algorithm adapts the Bellman equations to approximate the distribution of state-action values.
Instead of a parametrized function to learn the expectation of the return, they apply a parametrized function to estimate its distribution. Approximating the full distribution can provide more information and was shown to be advantageous in their implementation with increased robustness in the learning process \cite{Bellemare_2017}. 

Based on the usual Bellman equation (\ref{Eq_Bellman}) one can define a random return $Z$, with expectation equal to the state value function $\e [Z_t|S_t] = V(S_t)$ and derive a form of a distributional Bellman equation.
With $R(s)$ denoting the random variable reward, the recursive distributional equation is written as
\begin{equation}
Z_\pi(s) \overset{D}{=} R(s) + \gamma Z_\pi(s').
\end{equation}
The authors in \cite{Bellemare_2017} define $Z_\pi(s)$ as a mapping from the state to distributions over returns and prove for some distances defined over distributions, that the distributional Bellman operator can be written as a contraction operator that allows the use of repeated updates similar to the standard approach to learn the corresponding distribution.

They introduce an approximate distributional learning scheme that uses a parametric distribution, i.e a parametrized function $Z_\theta(\cdot)$, that can be updated by minimizing a cross-entropy term of the Kullback-Leibler divergence (KL-divergence) between the estimated distribution for the current state $Z_\theta(S)$ and a projection of the estimated distribution for the next state $Z_\theta(S')$ \cite{Bellemare_2017}.
 Alternative approaches utilizing for example the Gaussian case of approximate distributional learning can be found in \cite{Morimura_2010} as well as \cite{Tamar_2016}.
The distributional approach based on the parametric distribution presented in \cite{Bellemare_2017} has been implemented in other algorithms since, as for example \cite{Hessel_2018c} and was shown to positively impact sample efficiency and algorithmic performance in various applications.

The approach applied herein only depends on state values $V(S_t)$ and thus an algorithm to learn distributions of state values is presented below. It is an adjusted version of the parametric discrete distribution learning algorithm as introduced for state-action values in \cite{Bellemare_2017}, but reformulated herein for state values.

Let $N \in \mathbb{N}$ and define upper and lower bounds $V_{min},V_{max} \in \R$. Further, define a fixed set of atoms $z_i$ as
\begin{equation}
z_i = \{V_{min} + i \Delta z\,:0 \leq i < N\}, \nonumber
\end{equation} 
where the distances between atoms, i.e. the bin sizes, are given by:
\begin{equation}
\Delta z := \frac{V_{max}-V_{min}}{N-1}.
\end{equation}

These atoms serve to define the support $\{z_i\}$ of the estimated distribution and enable the estimation of $P(G_t = z_i|S)$ for each $z_j$ in the support $\{z_i\}$. 

The parametrized value distribution $Z_\theta(S)$ is then modelled by a parametrized function $f_{\theta}(S): \mathcal{S}  \rightarrow \R^N$, which maps\footnote{Generally, the output of the function can be standardized to reflect probabilities as $p_i(s) := \frac{e^{f_i(s)}}{\sum_j e^{f_j(s)}}$ or for example, if $f_\theta$ is a neural network, by applying a softmax activation to the output layer of the network.} the input state $S$ to an $N$-dimensional vector of estimated probabilities $p_i(s)$ for the atoms.

To enable the algorithm to exploit sample transitions to learn the parametrized distribution function, the authors \cite{Bellemare_2017} propose to compute a projection of the Bellman update for each atom $z_j$ on the support of $\{z_i\}$. The estimated probabilities of the next state are then distributed to its immediate members $m$ on the projected support.
Simplified, the projection is used to normalize the support such that estimates from the next step and current time step are in the same space. The projection accounts for the shift in the return by the next reward $r$ and the scaling introduced by $\gamma$. 
With $\hat{\mathcal{T}}$ being a projection operator, the projection of the Bellman update for each atom $z_j$ is given by:
\begin{equation}
\hat{\mathcal{T}}z_j := r+\gamma z_j. 
\end{equation}

Applying this projection and then distributing the estimated probabilities for the next state $p(s')$ to the nearest members of the projected support leads to a projected distribution $\Phi \hat{\mathcal{T}}Z_\theta(s')$ on the same support as the distribution of the current state.
Here, $\Phi$ is an operator symbol to clarify that one is looking at the distributed probabilities on the projection.

To clarify, let $b_j = \frac{\hat{\mathcal{T}}z_j -V_{min}}{\Delta z} $, $b_j \in [0,N-1]$ denote where the projected support falls relative to the atom $z_j$ of the support $\{z_i\}$. 
Then, the index of the nearest neighbouring supports are given by $l \leftarrow \lfloor b_j\rfloor, u \leftarrow \lceil b_j \rceil$ and the probabilities can be distributed to the nearest members $m_l,m_u$ proportional to their distances on the support given by $(u-b_j),(b_j -l)$.

With $[\cdot]_a^b$ bounding its argument in the range $[a,b]$, this leads to the following form of the $i$-th component:
\begin{align}
\Phi \hat{\mathcal{T}}Z_\theta(s')_i = \nonumber\\
\sum_{j=0}^{N-1} \left[1- \frac{|[\hat{\mathcal{T}}z_j]_{V_{min}}^{V_{max}}-z_i|}{\Delta z} \right]_0^1 p_j(s')
\end{align}

A sample loss that can be minimized by stochastic gradient descent is then given by the cross-entropy term of the Kullback-Leibler-divergence $D_{KL}\left(\Phi \hat{\mathcal{T}} Z_{\theta}(s') || Z_\theta(s)\right)$ and can be computed for a sample by:
\begin{equation}
\loss_{sample} = - \sum_i m_i \log p_i(s). \label{eq:DistLoss}
\end{equation}
This loss can be estimated by a computational friendly procedure as presented in \textit{Algorithm 1: Categorical Algorithm}  in \cite{Bellemare_2017}.

\section{The Novel Family of Models}\label{sec:approach}

The following section will first discuss the idea of the approach in general and convey the underlying motivation and advantages. Then, it addresses more specifically how such a model can be set up and provides a formal definition of its components.

\subsection{Concept and Motivation}
We combine aforementioned concepts from reinforcement learning and predictive knowledge to introduce the model approach. The first step is to simplify and realize that in the context of financial markets the value function itself is of interest and may play a pivotal role when it comes to the evaluation of assets. Instead of focussing on the more complex task of finding optimal decision rules, we take a step back and consider only the value functions for fixed policies. This gives rise to an approach that can be used to evaluate financial markets and compare different assets or serve as support to stabilize other machine learning approaches in finance.

Assume one knows the true value functions $v(\cdot)$, i.e. one knows the true expectation of the return $G_t=\sum_{k=t+1}^T \gamma^{k-t-1}r_k$ at any time step for individual assets and specific portfolios of interest. 
Then, at every time step $t$, one would not only have indication of how the markets will evolve, but could also compare different assets, anticipate changes and adjust holdings accordingly.

\begin{figure}[h]
  \centering
  \includegraphics[width=\linewidth]{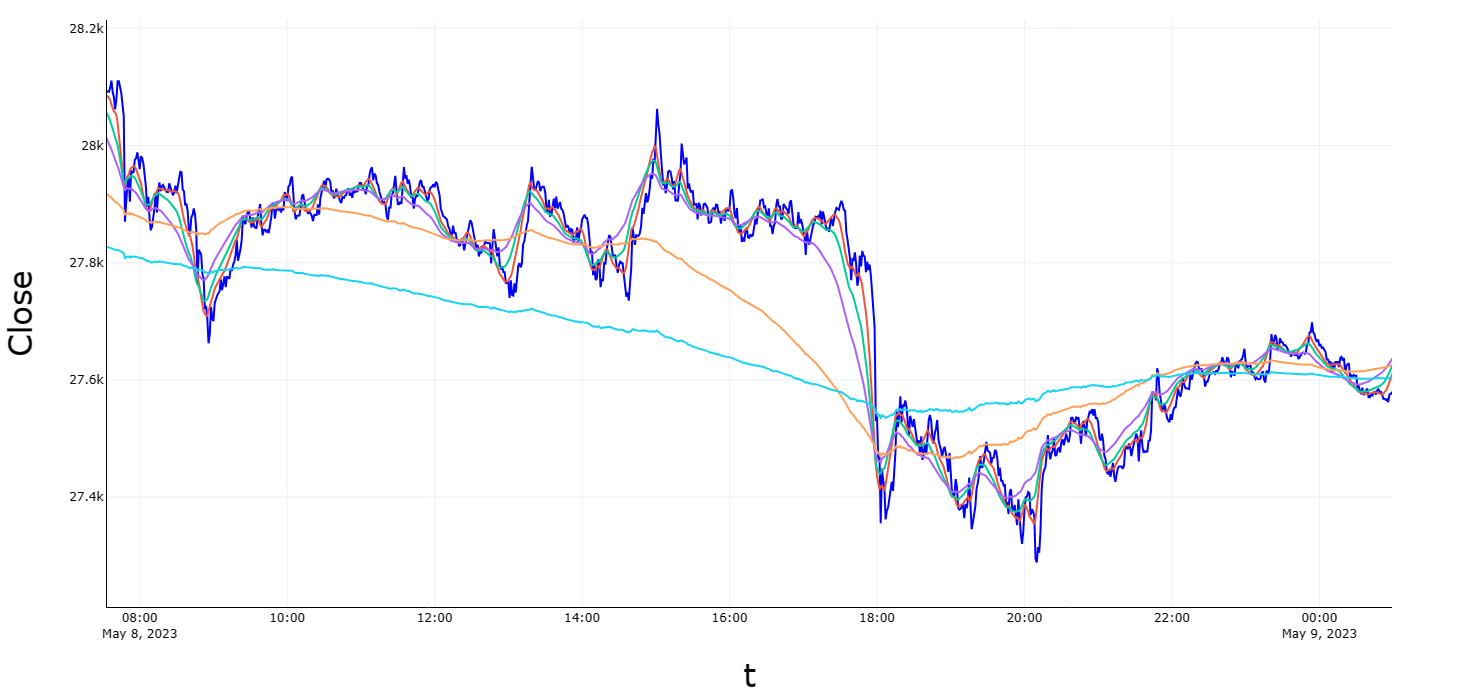}
  \caption{Realizations of return $G_t$ for different values of $\gamma$. Shown are the closing prices ($z_t$) of an asset in dark blue and prices combined with the realized returns for different values of $\gamma$. Specifically, the values plotted are $z_t\cdot e^{\tilde{G}_{t,\gamma_i}}$. Here, light blue is for the highest value of $\gamma$ ($\gamma=0.9975$) and the line in red represents the observed return for the lowest value ($\gamma= 0.8$).}
  \label{fig:figure_gamma_comparison}
\end{figure}

This is illustrated in Figure \ref{fig:figure_gamma_comparison}. It shows the true price of an asset (in dark blue) as well as the realized and observed returns $\tilde{G}$ combined with the prices for a specific choice of a reward definition (as introduced below). It is clearly visible that the return based lines are "leading" in time and imply future movement of the prices.

In reality of course one does not know these values ahead. But having good estimates of their expected values could be very beneficial and lead to better informed decisions and more optimal behaviour.\\

Drawing from the ideas of general knowledge and general value functions \cite{Sutton_2011} to stabilize the estimation and to force the model to better learn the underlying dynamics of financial markets, the proposed approach aims to not only learn the value function for one asset at a time, but to learn multiple value functions for different assets and portfolios in parallel. 
Considering multiple reward processes in the market and learning their value functions simultaneously should create not only a feature representation that contains rich information concerning the market, but also at the same time lead to a model that provides intuitive and comparable estimates for the future evolution of individual assets and portfolios.\\

There are two more aspects that need be considered. First, depending on the application, the focus of a model might be more on short-term or more on long-term dynamics. Second, estimating the expected return is very limited when it comes to supporting decisions and does not directly allow to include risk considerations or a more detailed analysis of the situation.

Therefore, the model is extended to take into account i) time considerations by simultaneously estimating state values for different time weightings $\gamma$, as well as ii) the more informative estimates of conditional distributions of the returns. 
These two extensions are now elaborated in detail.
\subsubsection*{i) Extension for time dependence}
The return is defined for a specific value of $\gamma$ and the expectation, i.e. the state value function, should be denoted by $V_{\gamma}$. 
The value $\gamma \in (0,1)$ represents the dependency on time and controls how much influence rewards at different future points in time have on the value function at time $t$.

Figure \ref{fig:figure_gamma_comparison}, where the prices weighted by observed realisations of a return are compared to the price for multiple values of $\gamma$, shows that the influence of future price movements, and thereby of future rewards increases for higher $\gamma$-values. 
The amount by which the realized line "moves ahead" of the true prices depends heavily on the choice of $\gamma$. 
The closer $\gamma$-values are to $1$, the earlier future price movements are reflected in the return. 
Thus, it might be beneficial to use higher or lower $\gamma$-values depending on the application of the model.
Further, in line with the idea of predictive knowledge, estimating the value functions for different choices of $\gamma$ in parallel, can intuitively force the feature representation learned by the model to focus more on short or long term dynamics and information.

Taking into account multiple values of $\gamma$ during the training process by including auxiliary tasks with different focus on time was shown to have a positive effect on stability and performance of reinforcement learning algorithms in \cite{Fedus_2019}. Thus, it can be expected to have a similar effect in the context at hand. \

Estimating the return at time $t$ for multiple different weightings of future changes has the additional benefit that at every time $t$, one not only has an estimation of future changes, but also information on how returns for different time horizons are estimated to relate to each other. A model that trustworthy estimates the correct values is an informative tool that offers more indication about future prices and greatly improves timing the market.

In an attempt to encourage both features and models with information regarding different time horizons, the model approach is thus extended to simultaneously estimate not only values for multiple strategies, but also for multiple $\gamma$-values.

\subsubsection*{ii) Extension to the distributional version}
The second adjustment concerns estimating the expected value.
The observed, realized values can vary from the expected value, and the expected value itself does not provide any information about the confidence of the model in its estimate, nor does it offer some risk related information.

Knowledge of the distribution of the return provides more information and if one learns the correct distribution for multiple time horizons for each time step $t$, one has most (if not all) the relevant information to take informed and optimal decisions. 
This naturally leads to the presented non parametric distributional version of the approach, where the state value function estimations are replaced by a distributional value function. 
Extending the initial approach to a distributional version is straightforward and leads to a model that estimates the distribution over the future returns for a given set of strategies and multiple $\gamma$-values.\\

Because these model use the \textbf{C}ontraction property to estimate expectations of the market return $\mathbf{G}$, we denote the approach using expected values as "\textbf{CG}-Model" (CGM) and the version that uses \textbf{C}ontraction to estimate the \textbf{D}istribution of the return $\mathbf{G}$ as "\textbf{CDG}-Model" (CDGM).

Both approaches are frameworks where one estimates in parallel multiple quantities for fixed strategies in financial markets using some value function approximation.

\subsection{Setup \& Components of the Model}
The general setup features an environment representing a financial market and an agent that interacts with it. This agent, or better called a valuation model, is equipped with a fixed set of deterministic policies. 
Each of these policies, called "base tasks" (or "base strategies") below, represents the process of following a clearly specified strategy. Any strategy or allocation, e.g. holding positions according to clear rules on how and when to allocate available funds over time, can be included. This can include investments in a single asset, portfolios of multiple assets and mixtures of asset classes. It is important to include strategies consisting of multiple assets such that the model can learn the dynamics under consideration of transaction costs and slippage.\\

Specifically, one simulates the interaction of the model with the environment and collects transitions $(S,\mathbf{r},S')$. In this setting, at every time step $t$, the model observes the new state of the environment and receives the rewards of every base task over the last period, where the rewards may be based on historical price data\footnote{Using the historical prices means one does not need to model the dynamics of the prices itself and can train a model free approach with transitions based on real world data. Companies that have real world transaction data from past strategies can exploit them to estimate value functions.}. If necessary, the agent then adjusts the allocations for each base task and moves on to observe the next state and reward at time $t+1$.
Based on the sampled transitions, one then estimates gradients of a loss function and uses gradient methods to train the model by minimizing the estimated loss. 

To stabilize the learning process of these models, prioritized replay buffers with priorities proportional to the total loss of a sample, target networks employing soft updates and n-step targets can be integrated.  \\

For the simulation of this process several components that make up the model, its training process and the environment need to be defined. 
Generally, for the model and training process a parametrized function and a loss function need to be defined. This function needs to be capable of approximating several value functions in parallel and the loss function is used to estimate gradients to perform updates to the function parameters.
In addition, a set of base tasks needs to be clearly specified. For the environment one needs to specify how states $S$ are represented as well as how the rewards for a given base task are calculated.

Each of these components and possible ways to specify them are now discussed. First, the model and its losses are formalized in Section \ref{sec:Model_learn} and some example base tasks are shown in Section \ref{sec:Base_task}. 
The components of the environment and possible reward definitions and their impact on the model are presented in Section \ref{sec:Env}.

\subsubsection{Model Function and Loss}\label{sec:Model_learn}
Formally, a set of $M$ base tasks $\{b_1,b_2,... b_M\}$ are defined, where each of these tasks $b_i,$ for $i = 1,2 \dotso , M$, represents a clearly defined strategy on how funds are allocated and adjusted over time.
Let the different $\gamma$-values be denoted by $\gamma_j$ for $j=1,2,\dotso,J$ and denote the respective state value functions by $V_{b_i,\gamma_j}(\cdot)=V_{i,j}(\cdot) \mapsto \mathbb{R}$ and distributional value functions by $Z_{b_i,\gamma_j}(\cdot)=Z_{i,j}(\cdot) \mapsto \mathbb{R}^{n_{atoms} \times 1}$, where the $n$-th atom of the estimated probability $Z_{i,j}$ is labelled $Z_{i,j,n}$.
Further, let $f(S,\theta)$ be some parametrized function depending on weight parameters $\theta$ with input representing the state $S_t$. For the CG-Model $$\hat{V}(\cdot)=f(S,\theta) \mapsto \mathbb{R}^{M \times J},$$
and for the CDG-Model $$\hat{Z}(\cdot)=f(S,\theta) \mapsto \mathbb{R}^{M \times J \times n_{atoms}}.$$
Let $\loss_{b_i,\gamma_j}=\loss_{i,j}$ denote the loss of the estimate for the $i-$th base task and $j-$th $\gamma$-value and define the total loss of the model $\loss_{total}$ as the average loss\footnote{Note: if different base tasks are of different importance, the total loss could also be defined as a weighted average.}
\begin{equation}
\loss_{total} = \frac{1}{MJ}\sum_{i=1}^M \sum_{j=1}^J \loss_{b_i,\gamma_j}.
\end{equation}

The loss for the individual base tasks depends on the choice of the value function approximation method. 
For the CG-Model the individual loss is defined by an appropriate distance measure\footnote{such as the mean squared error or a smooth Huber loss function.} between the value function at time $t$ and the recursive target at time $t+1$. 

For the current presentation, the loss for an individual base task $b_i$ and for a given $\gamma$-value $\gamma_j$ at time $t$ is estimated by:
\begin{equation}
\hat{\loss}_{i,j} = \left(\hat{V}_{i,j}(s;\theta) -  [r_{b_i,\gamma_j} +\gamma_j \hat{V}_{i,j}(s';\tilde{\theta})] \right)^2. \label{CG_loss}
\end{equation}

For the distributional version of the model, the loss is based on the KL-divergence
$D_{KL}(\Phi \hat{\mathcal{T}} Z_{\tilde{\theta}}(s') || Z_\theta(s))$ as presented above in Section \ref{sec:Distributional} and is estimated by 
\begin{equation}
\hat{\loss}_{i,j} = - \sum_{k}^{n_{atoms}} m_k \log p_k(s;\theta),
\end{equation}
where $p_k(s;\theta)$ is the $k-th$ element of the estimated distributions $\hat{Z}_{i,j}(s;\theta)$ and $m_k$ denotes the members to which the projected estimated probabilities are distributed. The probabilities for the next state are estimated as $Z_{i,j}(S';\tilde{\theta}) \in \mathbb{R}^{n_{atoms}}$ and given by the $(i,j)-th$ element of the output of $f(X_{t+1},\tilde{\theta}).$

\subsubsection{Example Base Tasks}\label{sec:Base_task}
A base task that is an allocation of funds to several assets can be modelled as a vector of the relative proportions to be invested in each asset and a rule on when adjust or rebalance the allocation. 

For example, in case of a fixed target allocation rebalancing could happen at every time step $t$, after a fixed time interval or whenever the current allocation deviates from a target allocation by more than some threshold.
A base task that rebalances periodically if the current allocation denoted by $\tilde{P}$ deviates too much from an initial target allocation $P_{target}=P_0$ due to price changes in underlying assets, is defined by the following policy $\pi_{b}(S)$
$$	\pi_{b}(S) = \begin{cases} P_0 & if \text{   }|\tilde{P} - P_0|>c,\, c \in \mathbb{R}\\ \tilde{P} & else. \end{cases}$$ 

An example of a set of base tasks of long-only strategies with $n$ different assets can be defined as follows. Let the first $n$ base tasks each represent the task of holding one specific asset $j$, i.e. $b_i= [ 0, \dotso, 1_{(j=i)},0, 0] \in \mathbb{R}^n$. Additionally, to allow the model to capture and estimate effects of transaction costs, expand the set by several base tasks of portfolios of multiple assets with clear defined rules for allocation adjustments. 
For illustration, define one base task as an equal weight portfolio $[\frac{1}{n},\frac{1}{n},\dotso, \frac{1}{n}]$ and specify additional base tasks of interest such as portfolios of a specific sector, industry or index. This can be expressed in matrix form.\\

There might exist some efficient or optimal set of base tasks to capture all relevant information and to have an optimal representation. But this is beyond the scope of this presentation\footnote{An interesting starting point regarding optimal geometrical representation might be \cite{Bellemare_2019}.}.

\subsection{Algorithm}
The full pseudo code for a CDG-Model is presented in \textit{Algorithm \ref{alg_cdgm}}.
Most steps apply analogously to the CG-Model by replacing the distributional losses and calculations in \textit{Algorithm \ref{alg_cdgm}} by the corresponding losses of the state value functions. 

Similarly, uniform sampling or on-line versions can be used by replacing the appropriate components in the learning procedure\footnote{I.e. for a uniform sampling replace the sample probabilities of transitions in the buffer to be same for all and drop importance sampling weights.}.

Note that for $M=1$, $J=1$ the models reduce to an estimate of value functions for a single strategy for a given time consideration.

\begin{algorithm*}[!h]
\caption{Pseudo-Code CDG-Model} 
\label{alg_cdgm} 
\begin{algorithmic} 
\STATE \hspace{2em}
\STATE
\STATE \textbf{DEFINITION AND INITIALIZATION}
\STATE \textbf{define} a set of base tasks $\{b_1,b_2,\dotso b_M\}$ for $M \in \mathbb{N}$ total number of tasks
\STATE \textbf{define} a set of discount factors $\{\gamma_1,\gamma_2,\dotso, \gamma_J \}$ for $J \in \mathbb{N}$ total number of $\gamma$
\STATE \textbf{define} number of atoms $n_{atoms}$, lower and upper bounds $V_{min},V_{max}$, fixed set of support values $\{z_i\}$
\STATE \textbf{define} updating speed $\tau$, learning rate $\alpha_l$, batch size $n_{batch}$
\STATE
\STATE \textbf{Initialize} a model function network $f_\theta(\cdot): \mathbb{R}^{dim(S)}\mapsto \mathbb{R}^{M \times J \times n_{atoms}} $ 
\STATE \textbf{Initialize} a target network as copy: $f_{\tilde{\theta},target}(\cdot)$, i.e. at initialization $\tilde{\theta}=\theta$
\STATE \textbf{Initialize} an empty replay buffer $\mathcal{B}_p,$ with parameters $\alpha_{\mathcal{B}},\beta_{\mathcal{B}}$ and maximum memory size of $n_{\mathcal{B},max}$
\STATE
\STATE \textbf{\RNum{1}: INTERACTION WITH ENVIRONMENT} 
\STATE collect and add transitions $(S,\mathbf{r},S')$ to the replay buffer $\mathcal{B}_p$ with default maximal priority $p_{max},$ 
\STATE where the rewards are 
\STATE \hspace{1em} i) for n-steps $=1$: $\mathbf{r} \in \mathbb{R}^M= [r_{b_1},r_{b_2},\dotso, r_{b_M}]$ 
\STATE \hspace{1em} ii) for n-steps $>1$: $\mathbf{r} \in \mathbb{R}^{M\times J}$, (i.e. for all $i,j:r_{t:t+n-1;b_i,\gamma_j},$ 
 see (\textit{Eq.} \ref{eq:r_in_nsteps})) and $S'=S_{t+n}$
\STATE
\STATE \textit{Once there are sufficient samples in the buffer, iterate between (\RNum{1}) interaction with the environment and (\RNum{2}) learning steps}
\STATE
\STATE \textbf{\RNum{2}: LEARNING STEP}
\STATE sample a batch of $n_{batch}$ experiences $\{e_1,\dotso,e_{n_{batch}}\}$ such that each experience $e_k$ in the buffer for $k \in 1,2,\dotso, n_{\mathcal{B}},$ where $n_{\mathcal{B}}\leq n_{\mathcal{B},max}$ (current number of samples in the buffer) is assigned sample probability $P(k) = p_{k}^{\alpha_{\mathcal{B}}}(\sum_j^{n_\mathcal{B}} p_j^{\alpha_{\mathcal{B}}})^{-1}$ and importance weight $w_k = \left(n_{\mathcal{B}} P(k) \right)^{-\beta_{\mathcal{B}}}$
\STATE 
\STATE \hspace{1em}\textbf{for each experience in batch} i.e. $e_i=(S,\mathbf{r},S')$ and corresponding importance weight $w_i$:
\STATE \hspace{3em} estimate probabilities for state $S$: $p(S;\theta)=f_\theta(S)$
\STATE \hspace{3em} estimate probabilities for next state $S'$ with target network i.e. $p(S';\tilde{\theta})= f_{\tilde{\theta},target}(S')$
\STATE
\STATE \hspace{3em} \textit{compute the distributional loss:}
\STATE \hspace{3em} \textbf{for each} $b_{i,\gamma_j}$ for $i=1,\dotso , M$, $j=1,\dotso , J$ \textbf{do}
\STATE \hspace{5em} set members $m_j = 0, j \in \{0, \dotso, (n_{atoms}-1)\}$
\STATE \hspace{5em} \textbf{for} $j \in 0, \dotso, (n_{atoms}-1)$ \textbf{do}
\STATE \hspace{7em} \textit{compute clipped projection:} 
\STATE \hspace{9em} $\hat{\mathcal{T}} z_j \leftarrow [r_{b_{i,\gamma_j}} + \gamma_j z_j]_{V_{min}}^{V_{max}} $ 
\STATE \hspace{7em} \textit{find relevant members and distribute probabilities:}
\STATE \hspace{9em} $b_j \leftarrow (\hat{\mathcal{T}}z_j -V_{min}) \Delta z $, $b_j \in [0,n_{atoms}-1]$
\STATE \hspace{9em} $ l \leftarrow \lfloor b_j\rfloor, u \leftarrow \lceil b_j \rceil$
\STATE \hspace{9em} $m_l \leftarrow m_l + p_j(S';\tilde{\theta})(u-b_j)$ 
\STATE \hspace{9em} $m_u \leftarrow m_u + p_j(S';\tilde{\theta})(b_j -l)$
\STATE \hspace{5em} \textbf{return} loss $\mathcal{L}_{b_{i,\gamma_j}} = - \sum_{j=0}^{n_{atoms}-1} m_j \log p_j(S;\theta)$
\STATE \hspace{3em} loss per base task $\mathcal{L}_{b_{i}} = \sum_{j=1}^J \mathcal{L}_{b_{i,\gamma_j}}$
\STATE \hspace{3em} \textbf{return} total loss per sample: $\mathcal{L}_{sample} = \sum_{i=1}^M \mathcal{L}_{b_{i}}$
\STATE 
\STATE \hspace{0em} \textbf{loss over batch:} $\mathcal{L}_{total} = \sum_{i=1}^{n_{batch}} w_i * \mathcal{L}_{sample_i}$
\STATE \hspace{0em} \textbf{calculate gradients} of total loss with respect to function parameters $\theta$ and perform a gradient step 
\STATE \hspace{0em} \textbf{update} priorities in replay buffer of the samples used in the batch, proportional to per sample loss, 
\STATE \hspace{3.8em} i.e. priorities $p_{sample} \propto \mathcal{L}_{sample}$
\STATE \hspace{0em}	\textbf{soft update} parameters of target network $\tilde{\theta} \leftarrow \theta \tau + (1-\tau) \tilde{\theta}$ 
\STATE \hspace{0em} \textbf{update} buffer parameter $\alpha_{\mathcal{B}},\beta_{\mathcal{B}}$ and learning rate $\alpha_{l}$

\end{algorithmic}
\end{algorithm*}

\subsection{Feature Creation \& Algorithmic Support}
Apart from being a potential valuation framework itself, the second benefit of the models is to use these concepts to enhance the feature creation or support and stabilize the learning process of other financial market algorithms. 
This is achieved by the framework as auxiliary losses to the learning process of the existing algorithms. 
The inclusion of distributional value functions or state value functions supports the feature creation by enforcing the learning of a 
more informative representation of the environment. This subsequently has great potential to improve the main algorithm or prediction functions. 
Further, even the estimated values, both distributions or expected values, can serve as (additional) inputs to trading- or prediction algorithms. 

\subsection{Components of the Environment}\label{sec:Env}
The components that make up the environment are subsequently discussed. Notice that while base tasks are an intrinsic part of the model, the states and rewards as part of the environment are external to the model.

\subsubsection{State Representation}
Generally, a state $S_t$ is defined by some set of information $I_t$ available at time $t$. This can be any data believed to be relevant to appropriately represent the state of a market at time $t$. Example of $I_t$ are prices of assets, order book information but also news, company financials or other constructed features. 
Importantly, the state $S_t$ does not need to be defined just as information $I_t$ (i.e. like a snapshot) at time $t$, but can be a collection of information over some number of last periods, i.e. $S_t = (I_t,I_{t-1},\dotso, I_{t-l})$, where $l$ represents the number of past time steps considered. Simplified, to make a decision today, everything that happened over the last week, or month is relevant.

Concerning the state space definition, the underlying Markov Assumptions $P(S_{t+1},r_{t+1}|S_t,S_{t-1},S_{t-2},\dotso) = P(S_{t+1},r_{t+1}|S_t)$ are relevant.
Importantly, the Markov Property is not an intrinsic property of the real world process but a property of the state space of the model of real processes. Any process can be modelled as a Markov Process \cite{Watkins_1989} by specifying the state space detailed enough to ensure that the current states capture all the relevant information needed to predict state transitions and rewards. 
Intuitively, if one includes all available information for all past periods, the Markov property holds. But since this is clearly not feasible, the relevant question becomes finding the appropriate set of information and past time periods to properly define the state $S_t$. 

For financial markets this is a very open and challenging topic and a proper set should be selected using statistical analysis, economic reasoning and careful investigation of the effects that different input data have on the models.
Even though above frameworks do not solve the dependency of any machine learning algorithm on its input data and the challenges in selecting optimal input data for financial models persist, the feature creation capabilities of these models are expected to lead to better representation for the same set of input.

\subsubsection{Reward Definition}

The definition of the rewards and their embedding into the environment is a modelling choice that allows to integrate and learn about any relevant quantities of interest in the trading process. 
These rewards can include possible costs and benefits\footnote{e.g. these might be transaction costs from rebalancing; slippage; possible fees \& rates of short selling or borrowing; coupon payments from bonds; dividends and so on.} of following a specified base task in addition to the change in the invested amount. The reward definition can also include considerations of risks that might be relevant for a true comparison of the performance of a strategy.
The general form of rewards and some explicit examples of rewards are explained in the following.\\

When evaluating an asset or an investment strategy arguably the most important quantity of interest is the change of the total value invested over time. 
To avoid any confusion between the "state value of a strategy" at time $t$ and the "total value invested in a strategy", the invested amount in a strategy at time $t$, is now called "worth" (or "wealth") of a strategy at time $t$ and is denoted by the random variable $W_t$ for a fixed base task.
As the change in $W_t$ over time is the quantity of interest, the rewards $r$ can generally be defined as the function
\begin{equation}
r_{t} = r(W_t,W_{t-1},W_{t-2},\cdots).
\end{equation}

The rewards can include risk considerations such as incorporating $\sigma_{W}$ or be expressed in terms of an utility function.

Herein, two simple versions of rewards defined solely as a function of the last two worth of the strategy, i.e. $r = r(W_t,W_{t-1})$, are shown. They are:
\begin{enumerate}[(i)]
\item log-returns: $r_{t} =  \log \left(\frac{W_t}{W_{t-1}}\right),$
\item cash-returns: $r_{t}  = W_t - W_{t-1}.$
\end{enumerate}

The incorporation of the reward as log-returns (i) into the Bellman updates from equation (\ref{Eq_Bellman}) is straight forward and state values remain comparable. However, using cash-returns (ii) suffers from the disadvantage that without relating to $W_{t-1}$, a direct comparisons in an economic sense of the numbers is not relevant. 
To remedy this, when learning the state value functions, one could fix the invested amount for each base task at the same level (which seems impractical) or adjust the updates to account for the dependence on the worth $W_t$.

Hence, to keep the estimated state values comparable, the updates in the case (ii) are slightly modified to learn the state value $V(S_t)$ at some time $t$ as the fraction of the worth, $\frac{V(s)}{W}$. The parametrized state value function is re-parametrized by factoring out $W_t$ explicitly,
\begin{equation}
\hat{v}_{\theta}(S_t) = f_{\theta}(S_t) \cdot W_t.
\end{equation}
 
The learned function $f_{\theta}(s)\mapsto\frac{V(s)}{w}$ estimates the return $G_t$ of one unit of cash invested in the strategy at time $t$. 
This allows to compare assets or strategies at different price levels.
It also reduces possible dependencies on the initial time step $t=0$, as every sample transition and its update depend only on the worth at time step $t$ and the reward stemming from investing that worth in the strategy. 
Note, that this does not modify the definition of the state value function or the recursive relationship. $W_t$ is a part of the state $S_t$ and the above formulation just re-parametrizes how the function is modelled.

The loss for a transition is then estimated by replacing $\hat{V}(\cdot)$ in (\textit{Eq.} \ref{CG_loss}) with the re-parametrization:
\begin{equation}
\loss = \left(f_{\theta}(S_t) \cdot w_t - r_t - \gamma[f_{\tilde{\theta}}(S_{t+1}) \cdot w_{t+1}]\right)^2. \nonumber
\end{equation}

Similarly, to utilize cash-returns (ii) in a distributional version, a minor adjustment to the updates in the learning process of the distributional version is needed.
Analogously to the re-parametrization above, an adjustment to the projection in the distributional case is proposed. With the aim to estimate distributions of a value, which multiplied by current worth is equal to the state value at time $t$, the space of the estimated distribution needs to be renormalized to account for the effect of the worth\footnote{In this slightly adjusted version, the model learns a distribution over the factor, such that 
$ \e [\hat{Z_t}W_t |S_t] = w_t \e [\hat{Z_t}|S_t]= w_t\sum_{i=1}^N p_i(s) z_i  = \hat{V}(S_t).$}. 
The following re-parametrized projection for the individual atoms is proposed:
\begin{equation}
\hat{\mathcal{T}}z_j := \frac{r_t}{w_t}+\gamma \frac{w_{t+1}}{w_t} z_j,
\end{equation} 
where $r_t,w_t,w_{t+1}$ are observed in the sample transitions. The estimated probabilities are then distributed to the immediate members of the projected support, and the same algorithm as above can be used to minimize the divergence.

\subsubsection{Definition of Worth}
The next relevant quantity is the evolution of the worth $W_t$ of a strategy over time. 
Generally, the worth of a base task at time $t$ depends on the costs of possible reallocations made at time $t-1$, the costs of following the base strategy over the period from $t-1$ to $t$, as well as the change in prices of assets considered in the strategy.

Any strategy of choice can be formulated and included in the model by defining the appropriate calculations to determine the worth of a strategy at time $t$. 
Two examples are introduced to illustrate this. One for holding a single asset and one for a long-only portfolio of multiple assets. They can be readily extended to include portfolios that are short in certain assets or hold multiple asset classes with different additional costs or similar. 

For the following examples we denote the price of an asset $i$ at time $t$ by $z_t^i$.

\subsubsection*{Example 1: Single-Asset}
For a simple base task consisting of holding one single asset, the change in worth is simply given by the change in asset prices and evolves according to $$W_{t+1}=  W_t  \frac{z_{t+1}}{z_{t}}.$$

\subsubsection*{Example 2: Long-only multiple asset allocation}

Consider a long-only asset allocation consisting of $N$ different assets defined as an $N$-dimensional vector of positions $P_t = [P_t^1,P_t^2,\dotso, P_t^N] \in \mathbb{R}^{N}$, such that $\sum_{i=1}^N P_t^i=1$.
The change in the strategy's worth can be defined as the cost from changing the positions to a new allocation plus the change in strategies value by the change in assets prices.

If at time $t$ the agent sets its position to $P_t$ and holds this position until time $t+1$, the position vector might change and fluctuate due to the changes in the prices of the involved asset. Hence, at the next decision time $t+1$, as part of the new state $S_{t+1}$, the position vector $P_t$ will have changed into an adjusted position vector, denoted by $\tilde{P}_t$.
Here, the subscript $t$ symbolizes the fact that the variable is the adjusted position\footnote{$P_t,\tilde{P}_t$ indicate the allocations dependence on the decision made at time $t$. Specifically, $P_t$ is set at time $t$ and $\tilde{P}_t$ is the result of $P_t$ evolving over the period starting at $t.$} relevant to how the allocation was set at time $t.$ 

At time $t$, the transaction costs occurring depend on the change from the current adjusted position $\tilde{P}_{t-1}$ to $P_t$. 
Let $\delta$ denote the proportional costs of changing a position, i.e transaction costs and possible slippage, then the direct cost of changing the position to $P_t$ at time $t$ is given by
$$cost_t = \delta\sum_{j=1}^N   |P_{t}^j - \tilde{P}^j_{t-1}|.$$ 
The change in the total amount invested from allocating the position at $t$ up to the new state $S_{t+1}$ is $ \left(\sum_{j=1}^N P_t^j \frac{z_{t+1}^j}{z_t^j} \right)$.
Combining the cost and change leads to the one-step relationship\footnote{The definition is inspired by similar definitions introduced in \cite{Moody_1997}, but adjusted to accredit costs to the decision made at time $t$. This more appropriately reflects the decision sequence of an MDP.} between strategies worth
\begin{equation}
W_{t+1}=  W_t (1 - \delta\sum_{j=1}^N | P_{t}^j - \tilde{P}_{t-1}^j|) \left(\sum_{i=1}^N P_t^i \frac{z_{t+1}^i}{z_t^i} \right). \nonumber
\end{equation}
Here, the $j$-th element of the adjusted position vector can be calculated using
\begin{equation}
\tilde{P}_{t}^j = \frac{P_t^j \frac{z_{t+1}^j}{z_t^j}}{\sum_{i=1}^N P_{t}^i \frac{z_{t+1}^i}{z_t^i}}. \nonumber 
\end{equation}

\subsection{Indicator}
The quantity $z_t\cdot e^{\tilde{G_t}}$  was shown in Figure \ref{fig:figure_gamma_comparison} for different values of $\gamma$. In a real application, the observed return $\tilde{G}$ can not be used at time $t$. However, it is simple to imagine possible indicators that could be constructed with good estimates of the expected return or the distribution.
For the reward case (i) the use of $z_t\cdot e^{\hat{V}_{t, \gamma_i}}$ is a natural choice. To make this point even clearer, consider the case for a single asset. There, $r_t =\log(\frac{W_{t+1}}{W_t}) = \log(\frac{z_{t+1}}{z_t})$ and
\begin{equation}
e^{\sum_{i=0}^T \gamma^i r_{t+1+i}} = \prod_{i=0}^T e^{\gamma^i \log\left(\frac{z_{t+1+i}}{z_{t+i}}\right)} = \prod_i^T \left(\frac{z_{t+1+i}}{z_{t+i}}\right)^{\gamma^i}. \nonumber
\end{equation}
In this product, $\lim_{i \rightarrow T} \left(\frac{z_{t+1+i}}{z_{t+i}}\right)^{\gamma^i} = 1$, i.e. the influence of future price changes gets smaller and vanishes eventually. Similarly for reward case (ii) $z_t \cdot (1+ f_\theta(S_t))$ can be used.

\section{First Results}
The following section presents some initial results from training above models on real financial market data. 
It serves as an initial proof of concept as well as discusses potential results and caveats, and illustrates properties and specific behaviour of the model. A difficulty in assessing the performance of the models lies in its high dimensionality and the presented distributions are just a subset of possibilities and cannot provide full coverage of the model properties.
The results are initial baselines which might be improved by optimizing different neural network structures, hyper-parameter settings and especially by considering more diverse sets of input data.

For all results, the models are tested on data that were not part of the training input. Typically, the last one or two month of available data are used for the test results and enough periods are discarded to avoid any overlap between training and testing data. The data sets consist of one minute prices and the states $S_t$ are created by combining returns based on close prices for multiple time windows for each of the involved assets. Specifically, for each asset, two time series based on returns and average close prices are constructed and the state is given by the concatenation of these series for all assets. Let $z_t$ denote the close price of an asset at time $t$ and $m_{t,l} = \frac{1}{l} \sum_{i=0}^l z_{t-i}$ the average of the close price over the last $l$ periods. The two time series used for a given asset are the quantities 
$\frac{z_t}{z_{t-l}}-1$, $\frac{z_t}{m_{t,l}}-1$ for a fixed set of lags\footnote{The used lags are $l=[1$, $2$, $20$, $30$, $45$, $60$, $90$, $120$, $180$, $240$, $360$, $720],$ as well as intervals of one minute bars of the following days equivalent $[1d,1.5d,2d,3d,5d,7d]$.} $l$.

For the quantitative studies performed herein, actual financial market data sets were exploited. In particular, the models were trained for three different asset classes using close prices of one minute intervals. The two asset classes stocks and ETFs comprise data\footnote{Specifically, split and dividend adjusted close prices for the following stocks: \textit{MSFT, AAPL, NVDA, AMZN, GOOG, TSLA, META, BRK.B, JNJ, PFE} and ETF: \textit{SPY, QQQ, IWM, GLD, DIA, XLF, XLE, USO, SLV, VXX, TLT}.} from \textit{First Rate Data}\footnote{\textit{www.firstratedata.com}} while close prices\footnote{For the following coins: \textit{BTCUSDT, ETHUSDT, BNBUSDT, NEOUSDT, LTCUSDT, ADAUSDT, XRPUSDT, EOSUSDT, IOTAUSDT, XLMUSDT}.} of crypto currencies were fetched from \textit{Binance}\footnote{\textit{www.binance.com}}.
For more information about the implementation see Appendix.

\subsection{Properties of the Distributions}

The estimated distribution for a time step $t$, is denoted by $\hat{D}_t$ and the estimated expected value $\hat{G}_t$ is calculated using $\hat{D}_t$ and the support values $z_i$ of the distribution. The realized and observed return will be denoted by $\tilde{G}_t$.
Evaluating $\hat{D}_t$ is difficult and a powerful statistical test might be hard to construct for numerous reasons. Mainly, because for each time step $t$, there is one estimated conditional distribution $\hat{D}_t$, estimating $P(G_t = z_i |S_t)$ for all support values $z_i, i \in \{ 1,2,\dotso n_{atoms} \}$, but only one realization of the observed return $\tilde{G}_t$. 
Thus, even if one knows or has estimated the correct distribution, the observed return as a random variable from that distribution could lie anywhere on the support and may not be close to the expected value nor have the same sign. Further, the true expected value or true underlying distribution can not be observed.
 
For a fixed time $t$, it is thus hard to make quantitative statements regarding the closeness of the estimated distribution to the true underlying distribution. 
Finding a meaningful and powerful statistical test combining all the different distributions and realizations presents a challenge because there are different distributions for every time step, each being the conditional distribution $P(G_t|S_t)$ given state $S_t$. One is left to look at average statistics over all time steps in the testing window, keeping in mind that they are neither identically distributed nor independent.
Nevertheless, it is important to share the results and discuss the models in view of a possible better interpretation and statistical analysis in the future.

The results in the following are thus of a more qualitative manner. 
Approaches to evaluate the distributions could be:
i) Count how often a realization falls into specific percentiles. This might indicate whether the distributions seem reasonable overall;
ii) The distribution of standardized statistics similar to z-scores might hint at existing bias;
iii) A substantial variation of the estimated distribution from $t$ to $t+1$ might point to inconsistencies, as one would expect some continuity, given $S_t$ and $S_{t+1}$ are very similar as they overlap in the included lags.

Hereafter, various observables are shown to illustrate different properties of the models and estimated distributions.

\subsubsection{Estimated distributions $\hat{D}_t$}

\begin{figure}[h]
  \centering
  \includegraphics[width=0.9\linewidth]{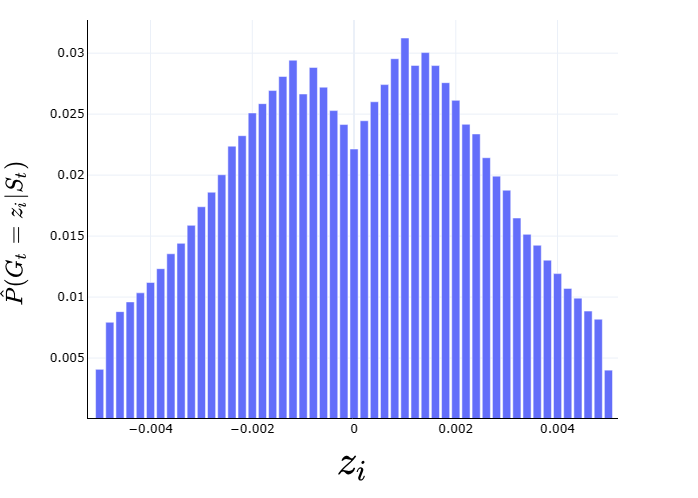}
  \caption{Estimated probabilities of the distribution $\hat{D}$ for some timestep $t$ in the test period and $\gamma=0.9$. Example of a bimodal distribution.}
  \label{fig:figure_multiple_ts_est_dist_bimodal}
\end{figure}

\begin{figure}[h]
  \centering
  \includegraphics[width=0.9\linewidth]{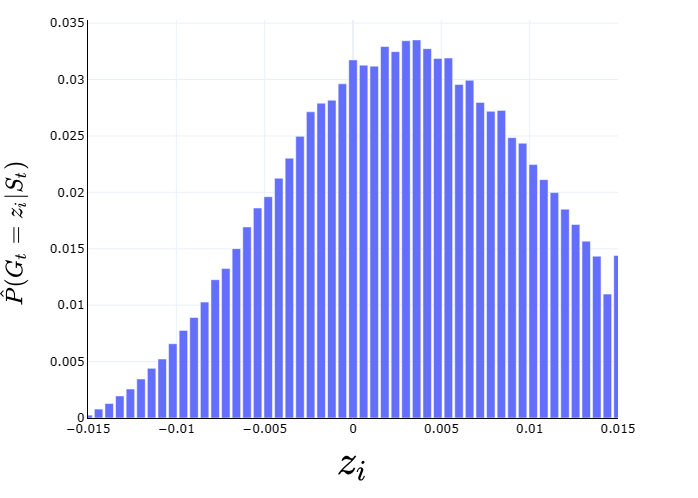}
  \caption{Estimated probabilities of the distribution $\hat{D}$ for some $t$ in the test period and $\gamma=0.9975$. Example of a strongly skewed distribution.}
  \label{fig:figure_skew_dist}
\end{figure}

\begin{figure}[!h]
  \centering
  \includegraphics[width=\linewidth]{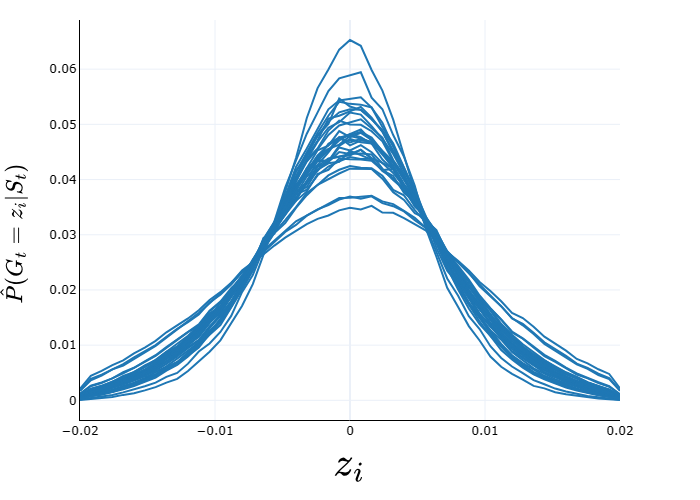}
  \caption{Estimated probabilities of the distribution $\hat{D}$ for multiple consecutive timesteps $t$ in the test period and $\gamma=0.8$. Example of a unimodal distribution.}
  \label{fig:figure_multiple_ts_est_dist_normalike}
\end{figure}

Over the testing periods most estimated distributions resemble some form of a Gaussian pattern, but also more bimodal or heavily skewed estimations emerge. Examples of estimated distributions $\hat{D}_t$ during the testing period are shown in Figures \ref{fig:figure_multiple_ts_est_dist_bimodal}, \ref{fig:figure_skew_dist} and \ref{fig:figure_multiple_ts_est_dist_normalike}. Here, bimodal, skewed and normal distributions are displayed as representative examples.

Figure \ref{fig:figure_multiple_ts_est_dist_normalike} shows the estimated distributions for a number of consecutive time steps and nicely illustrates a smooth change from rather peaked to a flatter estimation, possibly indicating different degrees of certainty of the model on how the values will likely evolve.

\subsubsection{Change of $\hat{D}_t$ with training progress}
\begin{figure}[h]
  \centering
  \includegraphics[width=\linewidth]{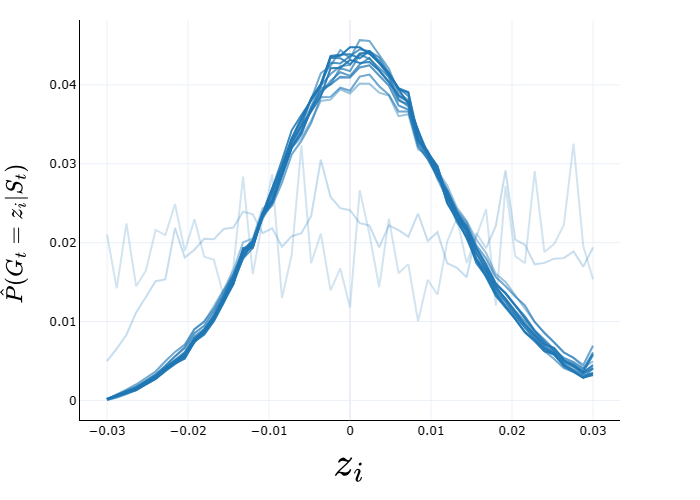} 
  \caption{Estimated distributions for the same time step $t$ at different steps in the learning progress, i.e. number of training epochs. The untrained model with random initialization of model weights is shown with lowest opacity. The longer the model was trained, the higher the opacity of the line representing the estimated distribution (blue line).}
  \label{fig:figure_est_dist_vs_ep}
\end{figure}

Figure \ref{fig:figure_est_dist_vs_ep} illustrates the changes in the estimates at some specific time $t$ in the test period depending on the number of epochs the model was trained for. 
The initially untrained and randomly initiated model predicts almost uniform probabilities. The more the model is trained, the more the estimation converges towards its final form.

\subsubsection{Estimated relation of $ \hat{G}$ to $\gamma$}

\begin{figure}[h]
  \centering
  \includegraphics[width=\linewidth]{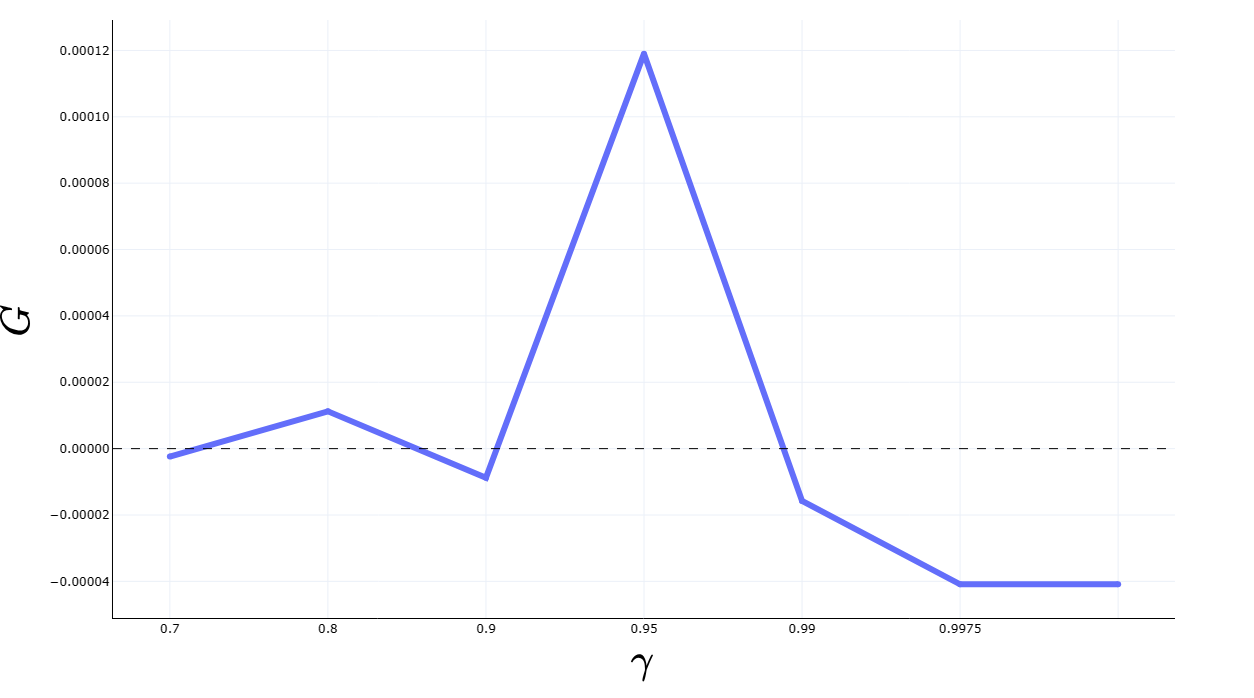} 
  \caption{The estimated expected values $\hat{\mathbb{E}}[G_t|S_t]$ (in blue) calculated with the estimated distributions $\hat{D}$ are shown against different $\gamma$-values (x-axis: $\gamma$, y-axis: possible values of $G_t$).}
  \label{fig:figure_gamma_line}
\end{figure}

The estimated expected values are shown in relation to the used values of $\gamma$ in Figure \ref{fig:figure_gamma_line} and Figure \ref{fig:comp_sb_nb_ggraph} shows examples with an overlay of the estimated distributions.
For trustworthy estimations these two graphical representations indicate how the models estimate the future evolution of the returns and might be indicative for timing the market.

One interesting take away from this might be the option to additionally robustify the models and estimation of the relation between expected values and $\gamma$-values. This might be achieved by creating additional auxiliary losses using the recursive definition of $G_t$ and the relation between $G_{\gamma_i}$ and $G_{\gamma_j}$, where $\gamma_i,\gamma_j$ are two different values. Specifically, additional losses exploiting the equality
\begin{equation}
G_{t,\gamma_i} - G_{t,\gamma_j} = \gamma_i G_{t+1,\gamma_i} - \gamma_j G_{t+1,\gamma_j}, \label{gline_aux_loss}
\end{equation} can force the models towards a more consistent and comparative estimation, as these quantities hold in the real world and should also hold in the estimated models. This aspect extends beyond the scope of this paper, but deserves future investigation.

\subsubsection{Comparison over different assets }
Even if the estimated distribution could not reliably predict the correct value, they still might be able to yield viable estimates of the relative strength or evolution of different assets and comparing $\hat{D}_t$ for different assets can provide further information.

\begin{figure*}[!ht]
  \centering
  \includegraphics[width=\linewidth]{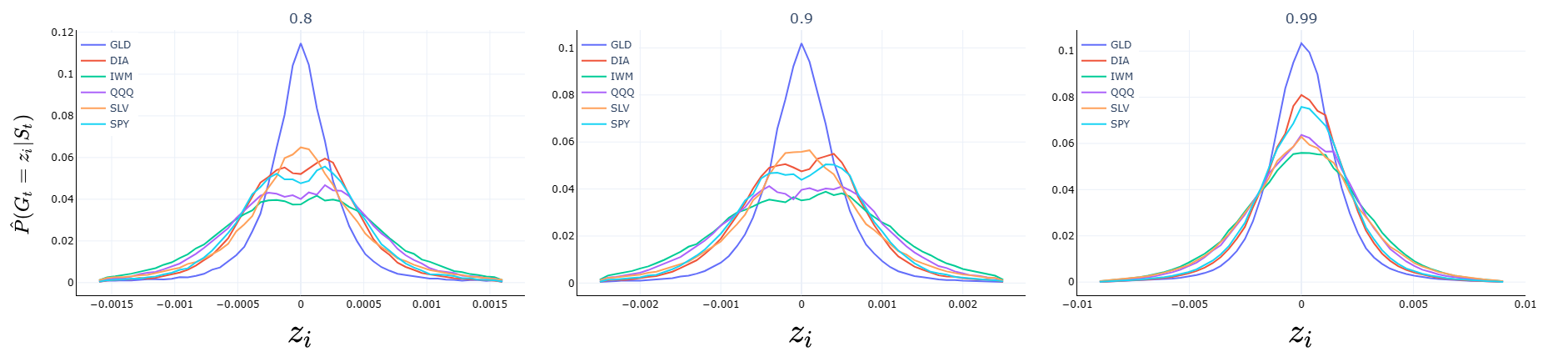}
  \caption{Estimated distribution $\hat{D}_t$ for multiple assets (in different colors) are shown for three different $\gamma$-values (left: $\gamma=0.8$, center: $\gamma=0.9$ and right: $\gamma=0.99$).}
  \label{fig:comparing_est_dist_cross_task}
\end{figure*}

\begin{figure}[h]
  \centering
  \includegraphics[width=\linewidth]{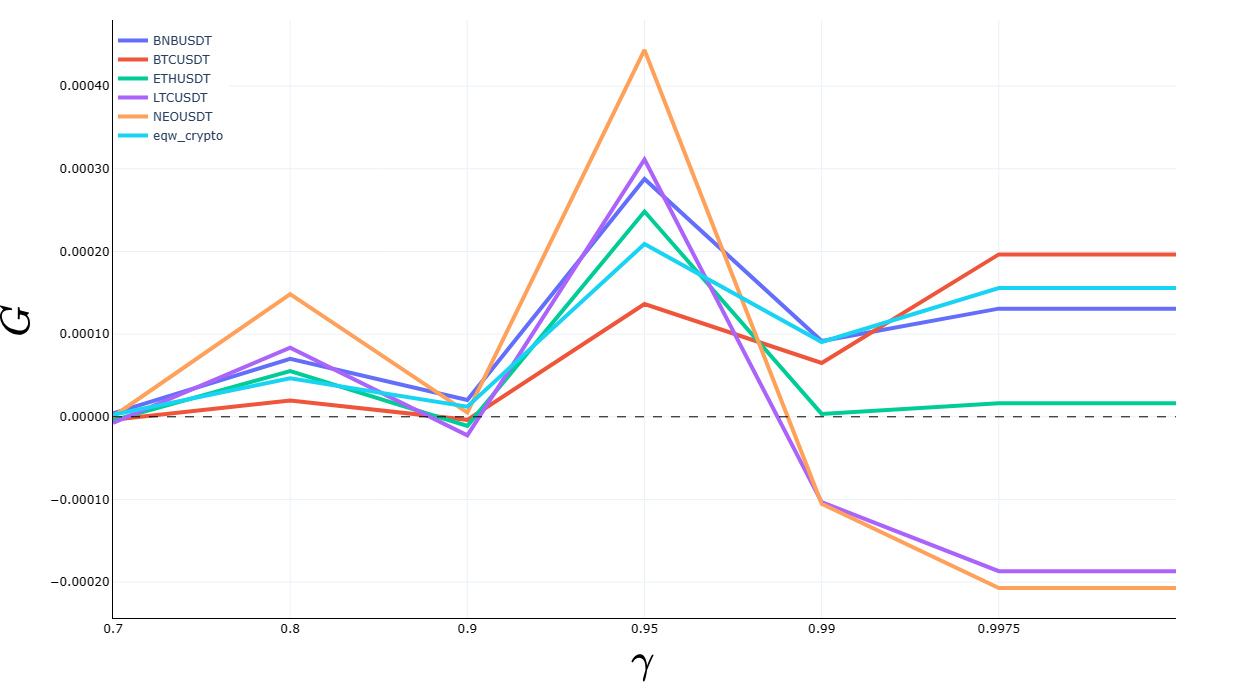}
  \caption{The estimated expected values $\hat{\mathbb{E}}[G_t|S_t]$ calculated with the estimated distributions $\hat{D}$ in relation to different $\gamma$-values are shown for multiple assets (in different colors).}  
  \label{fig:comparing_est_gline_cross_task}
\end{figure}

For illustration, Figure \ref{fig:comparing_est_dist_cross_task} shows $\hat{D}$ for different assets and choices of $\gamma$ and Figure \ref{fig:comparing_est_gline_cross_task} depicts the estimated relation between the expected value and $\gamma$ for different assets.
For trustworthy estimates, this could be valuable information to compare assets and aid asset selection, support the creation of pairs trading strategies or aid rebalancing decisions.

\subsection{Test-Statistics}
\begin{figure}[h]
  \centering
  \includegraphics[width=\linewidth]{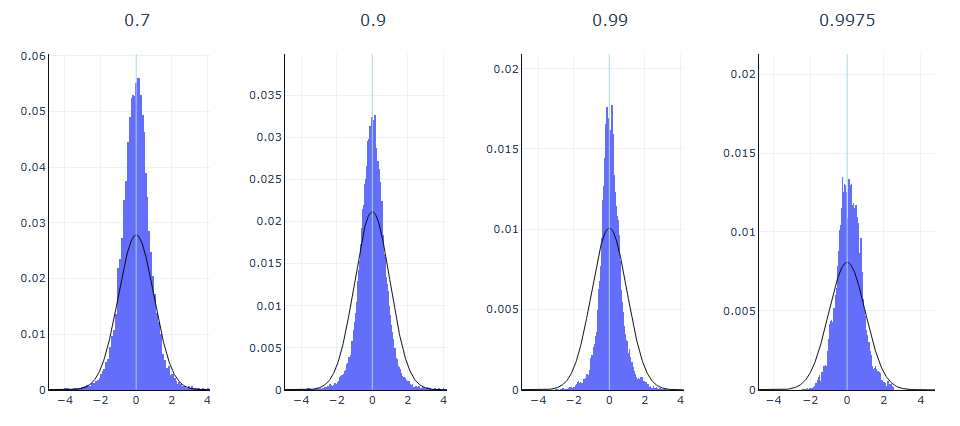}  
  \caption{Example how the distribution of the test statistic (in blue) compares to a normal distribution (normalized by bin-width, in black).}
  \label{fig:eg_z_scores}
\end{figure}

To investigate how the relationship between the estimated distribution, its expected value and the observed value behaves over the whole test set, two possible statistics are considered. First, a standardized statistic similar to z-scores is calculated and compared with a normal distribution to investigate a possible bias in the estimated expected values $\hat{G}$. Let $\tilde{z}_{t_j} = \frac{\hat{G}_{t_j}-\tilde{G}_{t_j}}{\sigma_{\hat{D}_{t_j}}}$ denote a standardized statistic and calculate this quantity for each $t_j$ in the test period. The histogram of these values is shown in Figure \ref{fig:eg_z_scores} for all time steps $t_j$ in the testing period. 
A normalized normal distribution is overlaid for comparison. Supposedly, over sufficient large test sample sizes, if there is no systematic bias in the expected values, one should expect the distribution\footnote{The expected form of the distribution of the test statistic is currently unclear and requires additional study outside the scope of this presentation.} of the test statistics to not reflect any clear patterns indicating larger or lower values.
Over the test cases conducted, the distribution of the test-statistic exhibits higher peaks, and the tails seem under represented compared to a normal distribution. The expected values seem to be generally larger than the observed values, with the effect being pronounced for larger $\gamma$-values. \\

\begin{figure}[!!h]
  \centering
  \includegraphics[width=\linewidth]{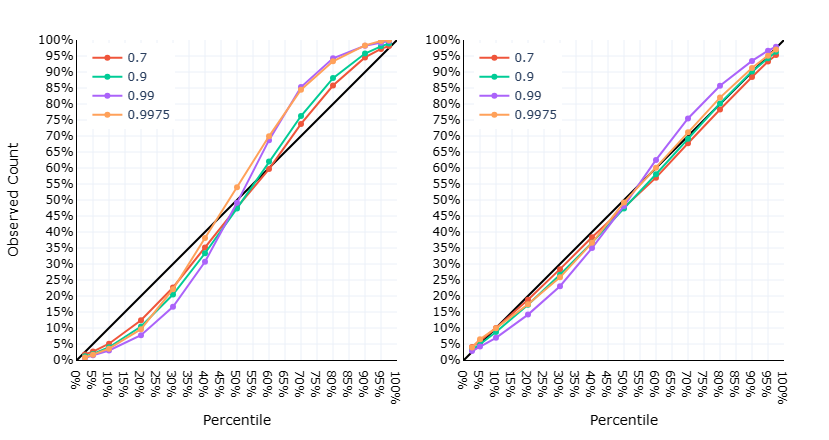}
  \caption{Normalized count statistics for different $\gamma$-values. Count of how often an observed value falls within a certain percentile of the estimated distributions normalized to the total number of testing steps. Plotted are the percentile values on the x-axis and the normalized counts on the y-axis. The black diagonal line indicates a perfect alignment (left: for a single asset, right: for equally weighted portfolio).}
  \label{fig:show_better_count}
\end{figure}

Another approach to assess the goodness of fit of $\hat{D}$ is to count how often the observed value falls within a certain percentile of the estimated distributions. If the estimated distributions are close to the true underlying distributions, one expects the counts to tend towards the percentile values as the number of testing steps goes to infinity. An example of such count statistic is shown in Figure \ref{fig:show_better_count} for a single asset as well as an equally weighted portfolio.

One interesting difference between count statistics for different base tasks of the model is apparent in Figure \ref{fig:show_better_count}. While percentile counts for single assets deviate more from the "ideal diagonal", base tasks of multiple assets such as the shown equally weighted portfolio seem to exhibit count statistics that mirror true percentile values much closer. It appears as if the models can better estimate the distributions for portfolios than single assets. This might be due to diversification effects in the portfolios resulting in lower uncertainty and thus better predictability. This agrees with our intuition, but deserves more thorough investigation.

\subsubsection{Influence of parameters}

To evaluate the influence of different parameters, a number of training runs were conducted by changing one of the parameters while keeping the others fixed. 
Specifically, cases with one, five or ten different $\gamma$-values and also one, five or ten considered assets, n-steps $\in \{1,5,10\}$ and the fixed support $z_i$ of the distribution with either small, medium or large values are tested. 

\begin{figure}[h]
  \centering
  \includegraphics[width=0.7\linewidth]{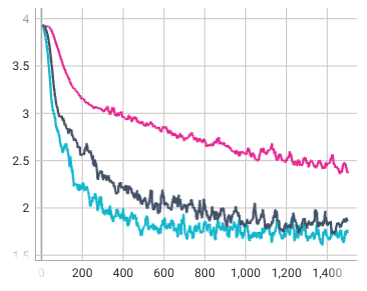}
  \caption{Variation of the average loss with the number of training epochs are shown for different choices of n-steps, with n-steps = 10 (light blue), n-steps=5 (dark blue), n-steps=1 (pink).
The number of training epochs are shown on the x-axis and the average loss on the y-axis (produced with Tensorflow \protect\cite{tensorflow2015_whitepaper}).\vspace{0pt}}
  \label{fig:tb_nsteps_comp}
\end{figure}

During training the only consistent observable pattern on the average loss manifested itself for different choices of n-steps as illustrated in Figure \ref{fig:tb_nsteps_comp}. The influence of the number of $\gamma$ and number of assets on the training error did not exhibit distinctive patterns. The average training error for larger values of n-steps $\in \{10,5\}$ led to faster decrease of the training error in the beginning and seems to reach lower values for higher number of trained epochs.

\subsubsection{Effect of support value size}
When considering different values for the support $\{z_i\}$, the use of larger absolute values for bounds $V_{\min},V_{\max}$ quickly leads to an estimation with all of the estimated probability on the inner-most support bins around 0. In contrast, for "smaller" or "medium" support values, the estimated probability is distributed over all bounds as it is expected.

\begin{figure}[!h]
  \centering
  \includegraphics[width=0.9\linewidth]{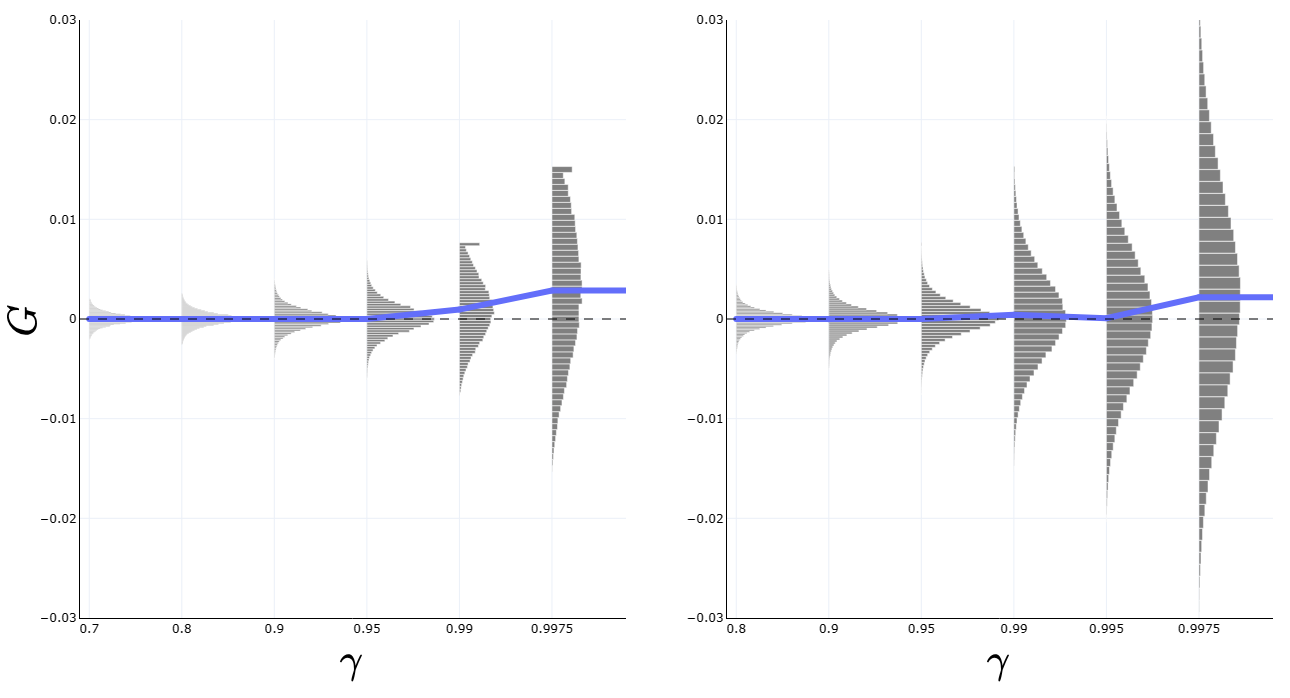}
  \caption{Estimated expected return for multiple $\gamma$-values using "smaller" bounds on the left vs. "medium" bounds on the right. In addition, the estimated distributions are overlaid as y-projection to illustrate the difference in used support values. The $\gamma$-values are shown on the x-axis and possible values for $G$ and the support values $\mathit{\{z_i\}}$ on the y-axis.\vspace{0pt}}
  \label{fig:comp_sb_nb_ggraph}
\end{figure}

\begin{figure}[!h]
  \centering
  \includegraphics[width=0.9\linewidth]{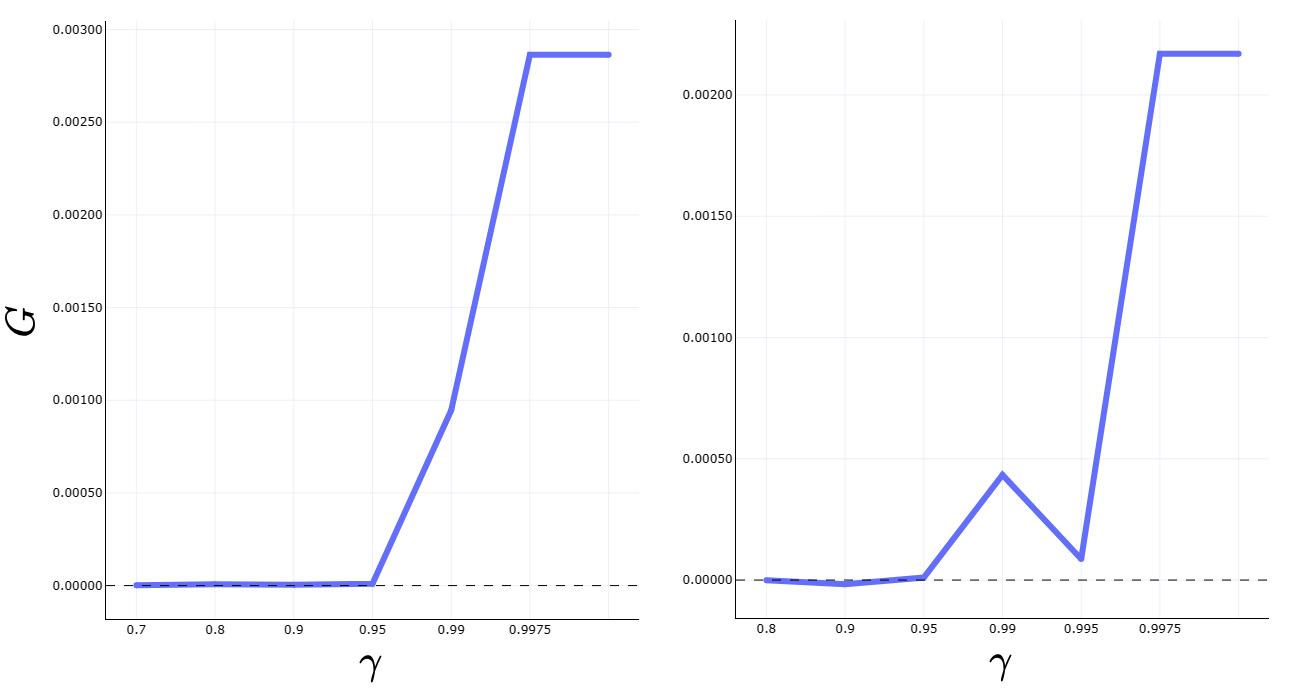}
  \caption{Estimated expected return for multiple $\gamma$-values using "smaller" bounds (on the left) and "medium" bounds (on the right).\vspace{0pt}}
  \label{fig:comb_gline}
\end{figure}

As the choice of support values and their influence on $\hat{D}_t$ might provide further information, Figures \ref{fig:comp_sb_nb_ggraph} and \ref{fig:comb_gline} show the results of two models trained with identical parameters apart from the choice of the support values. The support values are chosen to reflect a "smaller" case versus a "medium" support values case.

Both model settings estimate overall a very similar form of the relation between expected values and $\gamma$-values. The model with smaller support values seems to lose more of the distinction for smaller $\gamma$-values. However, the fact that both independently trained models converge to a very similar form for a specific point in time is an interesting observation. 

\section{Extensions \& Future Research}
Based on above discussions and findings, various possible paths for future research concerning these kinds of models may be chosen.
Besides the obvious hyper-parameter or neural network architecture optimization and diverse variations of input data, various directions can be explored. This can be the extension to other asset types like futures and options or portfolios containing short positions by introducing the appropriate reward functions.  
Further, one can investigate additional reward functions that utilize risk considerations and utility functions or even explore non-linear Bellman equations (see \cite{vanHasselt_2019}) by advancing from weighted sums to more involved functions.

It is also informative to consider relaxing the stationarity assumption by extending the framework to model non-stationarity or to partially observable Markov decision processes (POMDP \cite{Kaelbling_1998},\cite{Kimura_1997}).

A deeper study of the estimated distributions and how they relate to each other as well as to the distributions of asset prices might be interesting.

Other possible explorations include:

i) Investigations into the enhancement of existing trading algorithms by direct comparison of the performance and learning process of a trading algorithm with and without a) CDG-Model based auxiliary losses as well as b) including estimated distributions from a CDG-Model as input to an existing trading framework.

ii) Studies of optimal sets of base tasks.

iii) Focus on the estimation of the relation of expected values to gammas by including stabilizing losses as touched upon in (\textit{Eq.} \ref{gline_aux_loss}) or including $\gamma$ directly as an input to the approximation function.

iv) Trustworthy estimated distribution might be utilized to construct risk measures based on their characteristics such as volatility, skew or kurtosis as well as analogues of VaR and CVaR. It might even be possible to construct some portfolios similar to minimum variance portfolios based on estimated distributions.

v) The application of multiple distributional estimates for multiple time consideration might also be an interesting concept to incorporate into reinforcement learning algorithms outside the space of financial markets to support optimal decision.

\section{Conclusion}
The reinforcement learning return $G_t$ as a forward looking time weighted average over future rewards should be considered a quantity of high interest in the context of financial markets. Knowing the conditional distributions or at least reasonable estimates of them provides a competitive edge in financial markets, allowing for (near) optimal trading behaviour, both in asset selection as well as market timing.

In this research paper, a novel family of models that use "\textbf{C}ontraction to estimate \textbf{D}istributions of the return $\mathbf{G_t}$" (CDG-Model) was introduced, its motivation and advantages were described as well as challenges of assessing the model performance were discussed.

For completeness the relevant reinforcement learning theory and respective approximation problems and stabilization procedures were recalled. Based on these theoretical foundations the family of models and suitable training processes were introduced.
Further, a financial market environment and its components were defined, such that they are generally applicable to various assets classes and trading strategies. 
Two specific reward functions were defined and discussed, and examples for two base strategies, called base tasks, consisting of single assets or long-only portfolios were presented. 
These reward definitions and base tasks are just a small subset of all possible strategies and rewards that can potentially be evaluated and leveraged within this framework. 
A full pseudo code of the algorithm is provided in \textit{Algorithm \ref{alg_cdgm}}.\\
\vspace{-3pt}

The main advantages of the presented models over some point estimation process or mathematical model lie in their intuitive reasoning, the simplicity in assumptions as well as the underlying well studied properties of state value functions and distributional value functions.
If these assumptions hold, they guarantee uniqueness and existence of the state value functions. In addition, the flexible and intuitive framework allows to seamlessly integrate transaction costs and slippage into the evaluation and can be tailored to any asset class and trading strategy. \\
\vspace{-3pt}
The models are suitable to evaluate financial markets as well as to support feature creation and learning processes of machine learning based trading models. 
Specifically, they focus on the quantity return $G_t$ from reinforcement learning, defined as $G_t = \sum_{k=t+1}^T \gamma^{k-t-1}r_k$.
For a given state $S_t,$ representing the environment, the introduced family of models leverage either state value functions to estimate the expected return $\e[G_t|S_t]$ or distributional value functions to estimate the conditional probabilities $P(G_t|S_t)$ of the return for a fixed set of support values. Drawing from ideas of predictive knowledge to enhance the feature learning capabilities of the model and including time considerations, this estimation is performed in parallel for a set of multiple pre-defined, fixed base strategies and multiple different values for $\gamma$.

A particular difficulty in learning models in financial markets in general is to find and define suitable sets of input data. A further difficulty arises from the fact that these algorithms need to find a good feature representation, while simultaneously they have to solve some task like learning optimal behaviour or a prediction function. 
Some of these difficulties can be alleviated by applying the intuitive ideas from predictive knowledge such as estimating the distributions for multiple financial strategies in parallel. While the introduced models do not solve the dependency of any machine learning algorithm on its input data nor the challenges in selecting optimal input data for financial algorithms, they are expected to lead to better features for the same set of input.\\

\vspace{-5pt}
In order to provide a realistic context, the model framework was implemented and applied on real world prices of stocks, ETFs and crypto currencies. 
The obtained, initial results illustrate different forms of estimated distributions and give a first impression on the learning behaviour and model performance. They serve as proof of concept and provide first possible conclusions on the application of such models. 
The results show a certain consistency between independent training runs and exhibit smooth changes in estimated distributions over time.

One of the major challenges in the application lies in its high dimensionality and the difficulty to evaluate the estimated expected values and distributions. This is because there is only one realized observation for a given conditional estimation and test results are neither identically nor independently distributed. Creating meaningful and robust statistical tests is therefore difficult. Thus, two possible ways of assessing the quality of the estimated distributions such as percentile counts and a standardized statistic were discussed and applied to the test data. One interesting take away is that percentile counts exhibit better capabilities in estimating distributions for portfolios than for single assets. 

\vspace{5pt}
In conclusion, the CDG-Model family was introduced, its theory discussed and a successful proof of concept including first results from the application on real market data were provided. Various questions remain open and more robust statistical analyses are required. Still, the first lessons learned show that the models are capable of learning distributions that could very well represent realistic estimates and exhibit some consistency across independently trained models. 
Based on the model framework introduced and discussed herein, further studies are strongly encouraged both in pure scientific and in financial trading context.

\section*{Appendix: Implementation Details}\label{app:impl_details}

\begin{list}{\textbullet}{\leftmargin=0.5cm \itemindent=0pt}
\item \textit{Learning procedure: the learning process was done in an iterating fashion between i) generating and adding transitions to a prioritized replay buffer and ii) performing multiple learning steps by sampling batches of transitions. Transitions were generated by randomly initiating the environment at some time step $t$ in the training data and letting the agent and environment interact for sequences of at least $500$ steps to a maximum of $1000$ steps. Learning steps were only conducted once a sufficient amount of samples were collected in the replay buffer. All models were trained utilizing prioritized replay and target networks.}
\item \textit{Hyper-parameters setting: batch size of 512,
51 atoms, maximum buffer size of $80'000$ samples and first in first out fill procedure, ADAM optimizer \cite{Kingma2014adam} with initial learning rate of $0.00001$, soft target update parameter $\tau= 0.02$, 
buffer hyper-parameters: $\alpha_{\mathcal{B},0}= 0.75$, 
$\beta_{\mathcal{B},0}=0.25$, $\alpha_{\mathcal{B},end}= 0.0$, $\beta_{\mathcal{B},end}=1.0$,
with linear change over $2e4$ learning steps}
\item \textit{Choice of support values: support values $\{z_i\}$ were chosen in different sizes for different $\gamma$-values (smaller values for smaller $\gamma$-values). The same values for the support $\{z_i\}$ were used for stocks and ETFs, while different values were used for crypto currencies.}
\item Comment: hyper-parameters were based on typical numbers used in reinforcement learning research and were not optimized for the model training herein.
\end{list}

\section*{\textit{Acknowledgements}}
Many thanks for critical reading of the manuscript go to Prof. em. Dr. C. Grab (ETH Zurich).

\addcontentsline{toc}{section}{\protect\numberline{}{Bibliography}}
\bibliography{cg_research_paper_bibliography}

\end{document}